\documentclass[12pt, a4paper]{article}
 \usepackage[font=small,format=plain,labelfont=bf,up,textfont=normal,up,justification=justified,singlelinecheck=false]{caption}
\usepackage{subcaption}

\usepackage[a4paper, left=2cm,right=2cm]{geometry}
\usepackage[colorlinks=true,linkcolor=black,citecolor=teal,urlcolor=MidnightBlue,filecolor=black]{hyperref}
\usepackage{amsfonts}
\usepackage{amsmath,amssymb}
\usepackage{setspace}
\usepackage{slashed}
\usepackage[dvipsnames]{xcolor}
\definecolor{SchoolColor}{rgb}{0.6471, 0.1098, 0.1882} 

\usepackage[utf8,applemac]{inputenc}
\usepackage{tensor}
\usepackage{cite}
\usepackage{tikz}
\usepackage{graphicx}
\usepackage{bm} 

\usepackage[utf8,applemac]{inputenc}
\usepackage{tensor}
\usepackage{cite}
\usepackage{tikz}
\usepackage{graphicx}
\graphicspath{{figure/}}
\usepackage{array}
\usepackage{booktabs} 
\usepackage{multirow}

\usepackage{dcolumn}
\usepackage{bm}

\usepackage{verbatim}

\usepackage{textcomp} 
\usepackage{graphicx} 

\numberwithin{equation}{section}
\newcommand{\bea}{\begin{eqnarray}}
\newcommand{\eea}{\end{eqnarray}}
\newcommand{\be}{\begin{equation}}
\newcommand{\ee}{\end{equation}}
\def\nn{\nonumber}

\def\p{\partial}

\newcommand{\beqs}{\begin{eqnarray}}
\newcommand{\eeqs}{\end{eqnarray}}
\numberwithin{equation}{section}

\newcommand{\Rmnum}[1]{\uppercase\expandafter{\romannumeral #1\relax}}

\newcommand{\liu}{\color{red}}

\begin{document}
\begin{titlepage}

\begin{flushright}\vspace{-3cm}
{\small
\today }\end{flushright}
\vspace{0.5cm}
\begin{center}
	{{ \LARGE{\bf{Symmetry group at future null infinity \Rmnum{3}:
	
	Gravitational  theory}}}}\vspace{5mm}

	\centerline{\large{\bf Wen-Bin  Liu\footnote{liuwenbin0036@hust.edu.cn}, Jiang Long\footnote{
				longjiang@hust.edu.cn}}}
	\vspace{2mm}
	\normalsize
	\bigskip\medskip

	\textit{School of Physics, Huazhong University of Science and Technology, \\ Luoyu Road 1037, Wuhan, Hubei 430074, China
	}
	
	\vspace{25mm}
	
	\begin{abstract}
		\noindent
		{We reduce the gravitational theory in an asymptotically flat spacetime to future null infinity. We compute the Poincar\'e flux operators at future null infinity and construct the supertranslation and superrotation generators. The generators are shown to form a closed symmetry algebra by including a generalized gravitational duality operator. We could regard all the generators as the Hamiltonians with respect to the symmetry transformation in the boundary field theory. Our construction of the generators may relate to the BMS fluxes defined in the literature by adding counterterms to the Bondi mass and angular momentum aspects. }\end{abstract}
	

\end{center}

\end{titlepage}
\tableofcontents

\section{Introduction}

Recently, we proposed a new method to reduce the quantum field theory (QFT) in Minkowski spacetime to its future/past null infinity ($\mathcal{I}^+/\mathcal{I}^-$) \cite{Liu:2022mne,Liu:2023qtr}. In this method, we start with a massless bulk QFT whose fundamental fields are collected as $F_{\text{bulk}}$ and expand the bulk fields near $\mathcal{I}^+$ schematically as 
\be 
F_{\text{bulk}}= \frac{F(u,\Omega)}{r}+\sum_{k=2}^\infty \frac{F^{(k)}(u,\Omega)}{r^{k}}
\ee where $(u,r,\Omega)$ are the retarded coordinates of Minkowski spacetime. The field $F(u,\Omega)$ and higher order fields $F^{(2)}(u,\Omega),F^{(3)}(u,\Omega),\cdots$ become boundary fields of $\mathcal{I}^+$, and the bulk equation of motion (EOM) imposes constraints among the fields 
\be 
\mathcal{C}(F,F^{(k)})=0.
\ee 
The radiative modes of the bulk theory are encoded in the leading fall-off term $F(u,\Omega)$. 
After reducing the symplectic form to $\mathcal{I}^+$, the field $F(u,\Omega)$ and its time derivative $\dot{F}(u,\Omega)$ are conjugate variables and obey non-trivial commutation relations in the quantized theory. It is shown that the energy and momentum fluxes are completely determined by an energy flux density operator $T(u,\Omega)$ which is quadratic in $\dot F$
\be 
T(u,\Omega)\sim \dot{F}^2. 
\ee The smeared operator constructed from the energy  flux density operator 
\be \mathcal{T}_f=\int du d\Omega f(u,\Omega)T(u,\Omega)
\ee could form a higher dimensional Virasoro algebra with a divergent central charge. 
The central charge is proportional to the  number of propagating degrees of freedom of the bulk theory. When the test function $f(u,\Omega)$ is independent of the retarded time, the smeared operator $\mathcal{T}_f$ could be regarded as the generator of supertranslation. 

Similarly, the angular momentum and center-of-mass fluxes are determined by a flux density operator $M_A(u,\Omega)$ which is quadratic in  $F$ and $\dot F$ 
\be 
M_A(u,\Omega)\sim \dot F \nabla_A F-F\nabla_A\dot F.
\ee 
From the angular momentum flux density, one can define a smeared operator 
\bea
\mathcal{M}_Y=\int du d\Omega Y^A(u,\Omega)M_A(u,\Omega).
\eea When the test vector function $Y^A(u,\Omega)$ is independent of the retarded time, the smeared operator $\mathcal{M}_Y$ could be regarded as the generator of superrotation \cite{Barnich:2010eb, Barnich:2009se, Barnich:2010ojg, Barnich:2011mi, Campiglia:2014yka, Campiglia:2015yka}.  

In the scalar theory \cite{Liu:2022mne}, the smeared operators $\mathcal{T}_f$ and $\mathcal{M}_Y$  form a closed Lie algebra when $f(u,\Omega)$ is time dependent and $Y^A(\Omega)$ is not.  The closed Lie algebra is a direct generalization of the famous 
 Bondi-Metzner-Sachs (BMS) group \cite{Bondi:1962px, Sachs:1962wk,Sachs:1962zza} at future null infinity ($\mathcal{I}^+$) in asymptotically flat spacetime. When the central charge is zero, the closed Lie group could be regarded as a representation of the Carrollian diffeomorphism \cite{Ciambelli:2018xat,Ciambelli:2019lap} in the context of Carrollian manifold \cite{Une,Gupta1966OnAA,Henneaux:1979vn,Duval_2014a,Duval_2014b,Duval:2014uoa}.
 In the electromagnetic theory \cite{Liu:2023qtr}, one should introduce a new smeared operator which generates the generalized electromagnetic duality (EM duality) transformation to form an enlarged closed  algebra. The new operator could be interpreted as a helicity flux density operator. 

 Our method may provide new insight to flat space holography \cite{Strominger:2013jfa,Strominger:2017zoo,Kapec:2016jld,Pasterski:2016qvg,Pasterski:2017kqt,Raclariu:2021zjz,Pasterski:2021rjz,Donnay:2022aba,Donnay:2022wvx,Chen:2021xkw,Bagchi:2022emh,Chen:2023pqf,Saha:2023hsl} and to constructing more physically interesting Carrollian field theories \cite{Bagchi:2010zz,Bagchi:2016bcd,Bagchi:2019xfx,Bagchi:2019clu,Banerjee:2020qjj,Hao:2021urq,Henneaux:2021yzg,Bagchi:2022owq,Bagchi:2022xug,Bekaert:2022ipg,Rivera-Betancour:2022lkc,Schwarz:2022dqf,Dutta:2022vkg,Baiguera:2022lsw,Bekaert:2022oeh,Bagchi:2022eav,Saha:2022gjw}.  In this work, we will explore the boundary theory for Einstein gravity. We obtain a tensor field theory by projecting the linearized gravity to $\mathcal{I}^+$. The gravitational field has only two independent propagating degrees of freedom which are encoded in the symmetric traceless shear tensor $C_{AB}(u,\Omega)$.  
We find the energy and angular momentum flux density operators and define the supertranslation and superrotation generators, respectively. In order to make the definition of the superrotation generators sensible, one should generalize the Lie derivative variation to a covariant variation  which is compatible with the metric at $\mathcal{I}^+$ similar to what has been done in the electromagnetic theory. We also need to introduce a duality transformation operator to close the Lie algebra. The algebra turns out to be isomorphic to the one in the electromagnetic theory. The flux operators are shown to be equivalent to the Hamiltonians defined using the symplectic form of the boundary theory. We will also compare our construction of the flux operators with the BMS fluxes in the literature. 

The layout of this paper is  as follows. In section \ref{reviewform}, we will introduce the general framework and explain the terminology used in this paper. In section \ref{lineartheory}, we introduce the energy and angular momentum flux density operator in the linearized gravity theory. We also quantize the theory at $\mathcal{I}^+$ and find the supertranslation and superrotation generators. A closed Lie algebra is found by including a duality transformation operator $\mathcal{O}_g$. We compare the smeared operators with the BMS fluxes in the following section. We will summarize the results and discuss some further open questions in section \ref{cd}. Technical details are relegated to six appendices.

\section{Preliminaries} \label{reviewform}
In this section, we will introduce the general framework to obtain the boundary theory in Minkowski spacetime at future null infinity $\mathcal{I}^+$. 
\subsection{Boundary spacetime}
 In this work, the Minkowski spacetime $\mathbb{R}^{1,3}$ can be described in Cartesian coordinates $x^\mu=(t,x^i)$ 
 \bea 
 ds^2=-dt^2+dx^i dx^i=\eta_{\mu\nu}dx^\mu dx^\nu, 
 \eea where $\mu=0,1,2,3$ denotes the components of spacetime coordinates and $i=1,2,3$ labels the components of space coordinates. We also use the  retarded coordinate system $(u,r,\theta,\phi)$ 
 and write the Minkowski spacetime as 
 \bea 
 ds_{}^2=-du^2-2du dr+r^2\gamma_{AB}d\theta^Ad\theta^B,\quad A,B=1,2.
 \eea 
  The future null infinity $\mathcal{I}^+$ is a three dimensional Carrollian manifold 
  \bea 
  \mathcal{I}^+=\mathbb{R}\times S^2
  \eea with a degenerate metric 
\bea 
ds_{\mathcal{I}^+}^2\equiv\bm{\gamma}=\gamma_{AB}d\theta^Ad\theta^B.  \label{degemet}
\eea 
The spherical coordinates $\theta^A=(\theta,\phi)$ are used to describe the unit sphere whose metric reads explicitly as 
\bea 
\gamma_{AB}=\left(\begin{array}{cc}1&0\\0&\sin^2\theta\end{array}\right).
\eea 
We will also use the notation $\Omega=(\theta,\phi)$ to denote the spherical coordinates in the context. 
The covariant derivative $\nabla_A$ is  adapted to the metric $\gamma_{AB}$, while the covariant derivative $\nabla_\mu$ is adapted to the Minkowski metric in Cartesian frame. The integral measure on $\mathcal{I}^+$ is abbreviated as
\bea 
\int du d\Omega\equiv\int_{-\infty}^\infty du \int_{S^2}d\Omega,
\eea where the integral measure on $S^2$
is 
\bea 
\int d\Omega\equiv \int_{S^2}d\Omega=\int_0^\pi \sin\theta d\theta\int_0^{2\pi}d\phi.
\eea Besides the metric \eqref{degemet}, there is also a distinguished null vector 
\be 
\bm\chi=\partial_u
\ee which is to generate the retarded time direction. 

\begin{figure}
\centering
\hspace{0.8cm}
    \begin{tikzpicture}
        \filldraw[fill=gray!20,draw,thick] (0,4) node[above]{\footnotesize $i^+$} -- (2,2) node[above right]{\footnotesize $\mathcal{I}^+$} -- (4,0) node[right]{\footnotesize $i^0$}  -- (2,-2) node[below right]{\footnotesize $\mathcal{I}^-$} -- (0,-4) node[below]{\footnotesize $i^-$} -- cycle;
        \def\step{0.75};
        \foreach \i in {0,...,3}
        {\draw[very thin,blue] plot[smooth,tension=0.44] coordinates{(0,4) (0.6+\i*\step,0) (0,-4)};}
        \draw[latex-] (2.8,0.3) -- (6.6, 0.3);
        \node[fill=white,draw=black] at (6.6,0.3) {\footnotesize const. $r$ slices $\mathcal{H}_r$};
        \draw[latex-latex] (1.1,1.1) -- (0,0) -- (1.1,-1.1);
        \draw (1.1,1.1) -- (1.7,1.7);
        \draw (1.1,-1.1) -- (1.7,-1.7);
        \draw[latex-] (1.7,1.7) -- (6.6,1.7);
        \node[fill=white,draw=black] at (6.6,1.7) {\footnotesize $r$ increase, $u$ finite};
        \draw[latex-] (1.7,-1.7) -- (6.6,-1.7);
        \node[fill=white,draw=black] at (6.6,-1.7) {\footnotesize $r$ increase, $v$ finite};
        \draw[<->] (1.5,4) node[above left] {\footnotesize $u$} -- (2,3.5) -- (2.5,4) node[above right] {\footnotesize $v$};
    \end{tikzpicture}
    \caption{A series of constant $r$ hypersurfaces $\mathcal{H}_r$. As $r\to\infty$ while keeping $u$ finite, the slices approach to future null infinity $\mathcal{I}^+$. As $r\to\infty$ while keeping $v$ finite, the slices approach to past null infinity $\mathcal{I}^-$.}
    \label{sliceHr}
\end{figure}
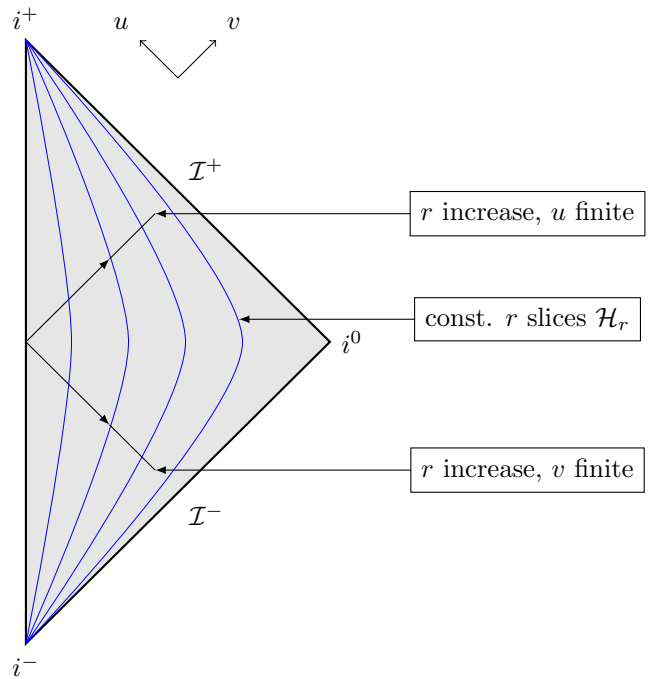

To obtain the metric of the Carrollian manifold \eqref{degemet} from bulk metric, one may choose a cutoff 
\bea 
r=R
\eea such that the induced metric on the hypersurface 
\bea 
\mathcal{H}_R=\{p\in \mathbb{R}^{1,3}|\ p=(u,r,\theta,\phi) \ \text{with}\ r=R\}
\eea is 
\bea 
ds^2=-du^2+R^2\gamma_{AB}d\theta^Ad\theta^B=R^2(-\frac{du^2}{R^2}+\gamma_{AB}d\theta^Ad\theta^B). \label{slice}
\eea The constant $r$ slices $\mathcal{H}_r$ are shown in figure \ref{sliceHr}. 
 We use a Weyl scaling to remove the conformal factor and  take the limit 
\bea 
R\to\infty
\eea while keeping the retarded time $u$ finite such that \eqref{slice} becomes the metric of the Carrollian manifold $\mathcal{I}^+$. We define the limit 
\bea 
\lim{}\hspace{-0.8mm}_+=\lim_{r\to\infty,\ u \ \text{finite}}\label{lim+}
\eea to send the quantities on $\mathcal{H}_r$ to $\mathcal{I}^+$. Similarly, taking the limit below 
\bea 
\lim{}\hspace{-0.8mm}_-=\lim_{r\to\infty,\ v\ \text{finite}}
\eea  sends the quantities on $\mathcal{H}_r$ to $\mathcal{I}^-$ where $v$ is the advanced time $v=t+r$.

\subsection{Boundary theory}
Consider a system which is described by a covariant Lagrangian $\mathcal{L}[F]$ in Minkowski spacetime $\mathbb{R}^{1,3}$, with the bulk field written as $F(t,\bm x)$.  The Lagrangian 4-form is defined as 
\be 
{\bf L}[F]=\mathcal{L}[F](d^4x),
\ee 
where $(d^4x)$ is the volume form of $\mathbb{R}^{1,3}$. 
A variation of the bulk field $F$ leads to 
\bea 
\delta{\bf L}[F]=\frac{\delta{\bf L}}{\delta F}\delta F-d\bm\Theta(F;\delta F).\label{variation}
\eea The first term on the right hand side is the equation of motion 
\be 
\frac{\delta{\bf L}}{\delta F}=0,
\ee which gives constraints on the bulk field. To solve the equation of motion, one should impose fall-off conditions for the field $F$ near $\mathcal{I}^+$
\bea 
F(t,\bm x)=\frac{F(u,\Omega)}{r}+\sum_{k=2}^\infty \frac{F^{(k)}(u,\Omega)}{r^k}.\label{falloff}
\eea
We have used the abbreviation 
\be 
F^{(1)}(u,\Omega)=F(u,\Omega)
\ee in the leading term of the fall-off conditions. The coefficients $F^{(k)}(u,\Omega),\ k\ge 1$ are called boundary fields corresponding to the bulk field $F$ since they are defined on the boundary manifold $\mathcal{I}^+$. By solving the bulk equation of motion asymptotically, one may obtain the solution space which is determined by the relations between the boundary fields $F^{(k)}(u,\Omega)$
\bea 
\mathcal{C}(F^{(k)})=0.\label{solspa}
\eea The second term on the right hand side of \eqref{variation} is exterior derivative of the presymplectic potential 3-form $\bm\Theta(F;\delta F)$. We have added a minus sign before this term by using the convention in \cite{2018arXiv180107064C} that the exterior derivative is anticommutative with the field variation, i.e., 
\be 
d\delta=-\delta d.
\ee The presymplectic form is defined as the variation of the presymplectic potential 
\bea 
\bm\omega(\delta F;\delta F)=\delta\bm\Theta(F;\delta F).
\eea The presymplectic form could be regarded as a 3-form in the spacetime and a 2-form in the phase space. The symplectic form of the theory is obtained by integrating the presymplectic form on a three dimensional hypersurface $\mathcal{H}$
\bea 
\bm\Omega^{\mathcal{H}}(\delta F;\delta F)=\int_{\mathcal{H}}\bm\omega(\delta F;\delta F)=\int_{\mathcal{H}} (d^3x)_\mu \ \omega^\mu(\delta F;\delta F).
\eea 
To find the symplectic form of the boundary theory at $\mathcal{I}^+$, we use the fall-off condition \eqref{falloff} and choose a series of hypersurfaces $\mathcal{H}_r$. By taking the limit $r\to\infty$ while keeping the retarded time $u$ finite, we find the symplectic form for the boundary theory 
\bea 
\bm\Omega(\delta F;\delta F)=\lim{}\hspace{-0.8mm}_+ \bm\Omega^{\mathcal{H}_r}(\delta F;\delta F).\label{bdysymp}
\eea 

The solution space together with the symplectic form \eqref{bdysymp} defines the classical theory on the Carrollian manifold $\mathcal{I}^+$. 
Interestingly, for the massless scalar theory and the electromagnetic theory, the symplectic form for the corresponding boundary theory can be written as 
\be 
\bm\Omega(\delta F;\delta F)=\int du d\Omega \delta F\wedge \delta \dot F. \label{symform}
\ee 
To quantize the boundary theory, one may define the Poisson brackets using the symplectic form and transform them into commutators. In order to get fundamental commutators, one can also use the mode expansion of the quantized field $F$ in the bulk and project it into the boundary in a suitable way. These two methods lead to the same commutators
\bea 
\ [F_I(u,\Omega),F_J(u',\Omega')]&=&\frac{i}{2}P_{IJ}\alpha(u-u')\delta(\Omega-\Omega'),\\ 
\ [F_I(u,\Omega),\dot{F}_J(u',\Omega')]&=&\frac{i}{2}P_{IJ}\delta(u-u')\delta(\Omega-\Omega'),\\ 
\ [\dot F_I(u,\Omega),\dot{F}_J(u',\Omega')]&=&\frac{i}{2}P_{IJ}\delta'(u-u')\delta(\Omega-\Omega')
\eea 
at $\mathcal{I}^+$ which have been checked in the scalar and vector theory.  The function $\alpha(u-u')$ is  defined through
\bea 
\alpha(u-u')=\frac{1}{2}[\theta(u'-u)-\theta(u-u')]\label{alpha}
\eea with $\theta(x)$ being the Heaviside step function, and the Dirac function on the sphere can be read out explicitly as
\be 
\delta(\Omega-\Omega')=\frac{1}{\sin\theta}\delta(\theta-\theta')\delta(\phi-\phi'). \label{diracsphere}
\ee We add a subscript in the field $F$ to represent its possible tensor structure. The tensor $P_{IJ}$ is symmetric under the exchange of $I$ and $J$
\be 
P_{IJ}=P_{JI}.
\ee It turns out to be 1 for the scalar field and  $\gamma_{AB}$ for the vector field.

\subsection{Leaky fluxes}
For any massless field theory, there could be bulk particles radiated to $\mathcal{I}^+$. Correspondingly, the Poincar\'e charges are sent to the boundary. These are called the leaky fluxes from bulk to boundary. The Poincar\'e current can be written concisely as 
\bea 
J^\mu=T^{\mu\nu}\xi_\nu 
\eea where $T^{\mu\nu}$ is the stress tensor of the bulk theory and $\xi^\mu$ is a Killing vector of Minkowski spacetime
\be 
\partial_\mu\xi_\nu+\partial_\nu\xi_\mu=0.\label{Keqn}
\ee There are ten independent Killing vectors  solving the Killing equation. 
For any global spacetime translation, the Killing vector  may be written as a superposition
\bea 
\bm\xi_c=\xi_c^\rho\partial_\rho,\qquad \xi_c^\rho=c^\rho
\eea where $c^\rho$ is any constant vector. Similarly, the Killing vector for any Lorentz rotation may be written as
\bea 
\bm\xi_\omega=\xi^\rho_\omega\partial_\rho,\qquad \xi_\omega^\rho=\omega^{\mu\nu}(x_\mu\delta^\rho_\nu-x_\nu\delta^\rho_\mu),
\eea where $\omega^{\mu\nu}$ is any constant antisymmetric tensor. 
The flux across a hypersurface $\mathcal{H}$ is defined as
\be 
Q^{\mathcal{H}}_{}=\int_{\mathcal{H}}(d^3x)_\mu J^\mu,\label{fluxdef}
\ee where $J^\mu$ is any conserved current satisfying $\partial_\mu J^\mu=0$. To obtain the Poincar\'e fluxes which are leaked to $\mathcal{I}^+$, we set $J^\mu=T^{\mu\nu}\xi_\nu$ and  choose a series of slices $\mathcal{H}_r$ to get
\bea 
Q^{\mathcal{H}_r}_{\bm\xi}=\int_{\mathcal{H}_r}(d^3x)_\mu T^{\mu\nu}\xi_\nu.\label{flux}
\eea  The normal covector of $\mathcal{H}_r$ is 
\bea 
dr \ \text{in retarded coordinates}=n_i dx^i\ \text{in Cartesian coordinates},
\eea where \be n^i=\frac{x^i}{r}=(\sin\theta\cos\phi,\sin\theta\sin\phi,\cos\theta)\ee  is the normal vector of $S^2$. We may define two null vectors in Cartesian coordinates 
\bea 
n^\mu=(1,n^i),\qquad \bar n^\mu=(-1,n^i).
\eea It is easy to show that
\bea 
n^2\equiv \eta_{\mu\nu}n^\mu n^\nu=0,\quad \bar{n}^2\equiv \eta_{\mu\nu}\bar{n}^\mu\bar{n}^\nu=0,\quad n\cdot\bar n\equiv\eta_{\mu\nu}n^\mu \bar{n}^\nu=2.
\eea Therefore, the normal vector of $\mathcal{H}_r$ can be written as 
\bea 
\frac{1}{2}(n^\mu+\bar{n}^\mu)\partial_\mu,
\eea and hence the flux \eqref{flux} across $\mathcal{H}_r$ becomes 
\bea 
Q^{\mathcal{H}_r}_{\bm \xi}=\frac{1}{2}\int_{\mathcal{H}_r}r^2 du d\Omega (n^\mu+\bar{n}^\mu)T_{\mu\nu}\xi^\nu.
\eea The integration measure on $S^2$ is defined through 
\be 
d\Omega=\sin\theta d\theta d\phi.
\ee Taking the limit \eqref{lim+}, we find the charge radiated to $\mathcal{I}^+$
\bea 
Q_{\bm\xi}=\lim{}\hspace{-0.8mm}_+\ \frac{1}{2}\int_{\mathcal{H}_r}r^2 du d\Omega (n^\mu+\bar{n}^\mu)T_{\mu\nu}\xi^\nu.\label{chargedef}
\eea In the following, we will show that the fluxes $Q_{\bm\xi}$ defined above are exactly those in \cite{Liu:2022mne,Liu:2023qtr}.
 \begin{enumerate}
     \item For a massless scalar theory with action 
     \bea 
     S=\int d^4x [-\frac{1}{2}\partial_\mu\Phi \partial^\mu \Phi-V(\Phi)],\label{scalaraction}
     \eea the stress tensor takes the form 
     \bea 
     T_{\mu\nu}=\partial_\mu\Phi\partial_\nu\Phi+\eta_{\mu\nu}\mathcal{L}(\Phi).\label{scalarstress}
     \eea The scalar field can be expanded asymptotically as 
     \bea 
     \Phi(t,\bm x)=\frac{\Sigma(u,\Omega)}{r}+\sum_{k=2}^\infty \frac{\Sigma^{(k)}(u,\Omega)}{r^k}.\label{falloffscalar}
     \eea In \cite{Liu:2022mne}, we have defined two smeared operators 
     \bea 
     \mathcal{T}_f&=&\int du d\Omega f(u,\Omega):\dot{\Sigma}^2:,\label{smT}\\ 
     \mathcal{M}_Y&=&\frac{1}{2}\int du d\Omega Y^A(u,\Omega)(:\dot\Sigma\nabla_A\Sigma-\Sigma\nabla_A\dot\Sigma:).\label{smM}
     \eea 
     \begin{itemize}
         \item When $\bm\xi=\partial_t$, we find 
         \bea 
         Q_{\partial_t}=\lim{}\hspace{-0.8mm}_+\int du d\Omega\ r^2 n^i T_{i0}=-\lim{}\hspace{-0.8mm}_+\int du d\Omega \ r^2 n_i T^{0i}.
         \eea This matches with the smeared operator \eqref{smT}
        with $f=-1$.
         \item When $\bm\xi=\partial_i$, we find 
         \bea 
         Q_{\partial_i}=\lim{}\hspace{-0.8mm}_+\int du d\Omega \ r^2 n_j T^{ji},
         \eea which is consistent with \eqref{smT} when $f=-n^i$.
         \item When $\bm\xi=x_i\partial_j-x_j\partial_i$, we find 
         \bea 
         Q_{x_i\partial_j-x_j\partial_i}=\lim{}\hspace{-0.8mm}_+\int du d\Omega\ r^2 n_k (T^{kj}x^i-T^{ki}x^j).
         \eea This is exactly the smeared operator \eqref{smM} with $Y^A=Y^A_{ij}$. The tensor $Y^A_{ij}$,  antisymmetric under exchanging $i,j$, denotes three Killing vectors on $S^2$, seeing Appendix \ref{ckvmink} for more details.
         \item When $\bm\xi=t\partial_i+x_i\partial_t$, we find 
         \bea 
         Q_{t\partial_i+x_i\partial_t}=\lim{}\hspace{-0.8mm}_+\int du d\Omega \ r^2 n_k (T^{ki}t-T^{k0}x^i).
         \eea One recovers the smeared operator $\mathcal{T}_f+\mathcal{M}_Y$ with $Y^A=Y^A_i$ and $f=\frac{1}{2}u\nabla_A Y^A_i$, where the tensor $Y^A_i$ denotes three strictly conformal Killing vectors on $S^2$, defined in  Appendix \ref{ckvmink}.
     \end{itemize}
     \item For the electromagnetic theory, 
     the fall-off conditions are 
     \bea  
a_u(t,\bm x)&=&\frac{A_u(u,\Omega)}{r}+\mathcal{O}\left(\frac{1}{r^2}\right),\label{fallau}\\
a_A(t,\bm x)&=&A_A(u,\Omega)+\frac{A_A^{(1)}(u,\Omega)}{r}+\mathcal{O}\left(\frac{1}{r^2}\right)\label{fallaA}
\eea for the vector potential $a_\mu$ in the radial gauge $a_r=0$.
     One can also reproduce the smeared operators defined in \cite{Liu:2023qtr} similarly. Note that there is an additional conserved current  $j_{\text{duality}}^\mu$ that is related to the electromagnetic duality transformation for free Maxwell theory. Therefore, one may choose the EM duality current 
     in \eqref{fluxdef} and find 
     \bea 
     Q_{\text{duality}}=\lim{}\hspace{-0.8mm}_+\ \frac{1}{2}\int_{\mathcal{H}_r}r^2 du d\Omega (n^\mu+\bar{n}^\mu)j^\mu_{\text{duality}}=\int dud\Omega \dot A^C A^B\epsilon_{BC}
     \eea 
     which may be weighted by a parameter $g(\Omega)$. Thus,  after taking normal order, the operator 
     \begin{align}
         \mathcal{O}_g=\int dud\Omega g(\Omega):\dot A^C A^B:\epsilon_{BC}
     \end{align}
     is exactly the generalized EM duality operator we have found.
 \end{enumerate}
 
 \subsection{Hamiltonians from boundary theory}
In this subsection, we will define the Hamiltonians using the symplectic form of boundary theory defined on the Carrollian manifold $\mathcal{I}^+$. One may define an infinitesimal  Carrollian diffeomorphism generated by a vector $\bm \xi$ through \cite{Liu:2022mne}
\bea 
\mathcal{L}_{\bm\xi}\bm\chi=\mu \bm\chi, \label{cdiff}
\eea where $\mathcal{L}_{\bm\xi}$ is the Lie-derivative along the direction of $\bm\xi$. 
The general solution of \eqref{cdiff} is 
\bea 
\bm\xi=Y^A(\Omega)\partial_A+f(u,\Omega)\partial_u.
\eea Note that the vector field $Y^A(\Omega)$ is time independent in Carrollian diffeomorphism. When $Y^A$ is time dependent, it violates the definition \eqref{cdiff}
 and breaks the null structure of $\mathcal{I}^+$. It has been shown that the Carrollian diffeomorphism is a physical transformation which corresponds to the radiation flux  from bulk to boundary \cite{Liu:2022mne}. Therefore, 
we may define a Hamiltonian $H_{\bm\xi}$ \cite{Iyer:1994ys,Wald:1993nt,Wald:1999wa} whose infinitesimal variation is 
\bea 
\delta H_{\bm\xi}=i_{\bm\xi}\bm\Omega(\delta F;\delta F)\label{vacharge}
\eea where $i_{\bm\xi}$ is the interior product in the phase space. More explicitly, we have
\bea 
i_{\bm\xi}=\delta_{\bm \xi}F\frac{\partial}{\partial\delta F}\quad\Rightarrow \quad i_{\bm\xi}\delta F=\delta_{\bm\xi}F.\label{interiorprod}
\eea 
Substituting the symplectic form \eqref{symform} into \eqref{vacharge}, we find the variation of the Hamiltonian $H_{\bm\xi}$ corresponding to the vector field $\bm\xi$ 
\bea 
\delta H_{\bm\xi}=2\int du d\Omega \delta_{\bm\xi}F \delta\dot F.\label{canocharge}
\eea We will explore this formula for scalar theory and electromagnetic theory in the following. 

\subsubsection*{Scalar theory}
 For the previous scalar theory, we have $F=\Sigma$. The Carrollian diffeomorphism may be split into two parts 
    \bea 
    \bm \xi=\bm \xi_f+\bm\xi_Y, 
    \eea where 
    \bea \bm\xi_f=f(u,\Omega)\partial_u,\quad \bm\xi_Y=Y^A(\Omega)\partial_A
    \eea and the scalar transforms as\footnote{As a matter of fact, the transformation law $\delta_{f/Y}\Sigma$ is induced from the bulk BMS transformation. We believe there is an intrinsic way to derive this law at boundary which will be our future interest.}
    \bea 
    \delta_f \Sigma&=&f(u,\Omega)\dot\Sigma,\\ 
    \delta_Y \Sigma&=&\frac{1}{2}u\nabla_A Y^A(\Omega) \dot\Sigma+Y^A(\Omega)\nabla_A\Sigma+\frac{1}{2}\nabla_A Y^A(\Omega) \Sigma.
    \eea 
    For $\bm\xi=\bm\xi_f$,
    we find 
    \bea 
   \delta H_f=2\int du d\Omega f(u,\Omega) \delta \dot\Sigma \dot\Sigma=\delta \int du d\Omega f(u,\Omega)\dot\Sigma^2.
    \eea Therefore, there is a natural integrable flux in the  boundary theory 
    \bea 
    H_f=\int du d\Omega f(u,\Omega)\dot\Sigma^2.
    \eea This is exactly the smeared operator $\mathcal{T}_f$.
    When $\bm\xi=\bm\xi_Y$, we find 
    \bea 
    \delta H_Y&=&2\int du d\Omega \delta \dot \Sigma (Y^A\nabla_A\Sigma+\frac{1}{2}\nabla_A Y^A \Sigma+\frac{1}{2}u\nabla_A Y^A \dot\Sigma)\nn\\
    &=&\delta \int du d\Omega \dot\Sigma (Y^A\nabla_A\Sigma+\frac{1}{2}\nabla_A Y^A\Sigma+\frac{1}{2}u\nabla_A Y^A \dot\Sigma).
    \eea We have used integration by parts at the second step.  Therefore, we find the Hamiltonian corresponding to  $\bm\xi_Y$ 
    \bea 
    H_Y&=&\int du d\Omega \dot \Sigma (Y^A\nabla_A\Sigma+\frac{1}{2}\nabla_A Y^A\Sigma+\frac{1}{2}u\nabla_A Y^A \dot\Sigma)\nn\\
    &=&\mathcal{M}_Y+\mathcal{T}_{f=\frac{1}{2}u\nabla\cdot Y}.\label{hymy}
    \eea 
    Once subtracting the second part, it becomes the smeared operator $\mathcal{M}_Y$.
    
\subsubsection*{Electromagnetic theory}
In the electromagnetic theory, we have $F=A_A$ and  the fundamental field is $A_A$ whose variation under $\bm\xi_f$ reads  
    \bea 
    \delta_f A_A&=&f(u,\Omega)\dot{A}_A.
    \eea Now it is straightforward to find 
    \bea 
   \delta H_f=\delta\int du d\Omega f(u,\Omega)\dot{A}_A\dot{A}^A\quad\Rightarrow\quad H_f=\mathcal{T}_f.
    \eea 
    When $\bm\xi=\bm\xi_Y$, as has been shown in \cite{Liu:2023qtr}, we may replace the variation $\delta_Y A_A$ to covariant variation $\delta\hspace{-6pt}\slash_Y A_A$
    \bea 
\delta\hspace{-6pt}\slash_YA_A=\frac{1}{2}u\nabla_A Y^A \dot A_A+Y^C\nabla_C A_A+A_C\nabla_AY^C-\frac{1}{2}\Theta_{AC}(Y)A^C
    \eea where the symmetric traceless tensor $\Theta_{AB}(Y)$ is
    \bea 
    \Theta_{AB}(Y)=\nabla_A Y_B+\nabla_B Y_A-\gamma_{AB}\nabla_C Y^C.\label{Thdef}
    \eea In this case, we may modify the equation  \eqref{interiorprod} to 
    \bea 
   i_{\bm\xi}=\delta\hspace{-6pt}\slash_{\bm\xi} F\frac{\partial}{\partial\delta F}\quad\Rightarrow\quad i_{\bm\xi}\delta F=\delta\hspace{-6pt}\slash_{\bm\xi}F.\label{modiinter}
    \eea Now it is straightforward to find 
    \bea 
    \delta H_Y=\delta (\int du d\Omega \dot{A}^A \delta\hspace{-6pt}\slash_{Y}A_A).
    \eea Therefore, we get the corresponding flux
    \bea 
    H_Y=\int du d\Omega \dot{A}^A \delta\hspace{-6pt}\slash_{Y}A_A=\mathcal{M}_Y+\mathcal{T}_{f=\frac{1}{2}u\nabla\cdot Y}.
    \eea 
    
    There is a new operator corresponding to electromagnetic duality transformation in the free electromagnetic theory \cite{Liu:2023qtr}. We could not find a vector field in spacetime for this transformation. However, we can find the field variation due to the generalized EM duality transformations with parameter $g(\Omega)$
    \bea 
    \delta_g A_A=-g(\Omega)\epsilon_{AB}A^B(u,\Omega).
    \eea The infinitesimal variation of the  corresponding Hamiltonian reads
    \bea 
    \delta H_g&=&-\int du d\Omega g(\Omega)(\delta\dot A^A  \epsilon_{AB}A^B-\delta A_A  \epsilon_{AB}\dot{A}^B)\nn\\
    &=&\delta \int du d\Omega g(\Omega) \dot{A}^A\epsilon_{BA}A^B.
    \eea Therefore, the  Hamiltonian takes the form 
    \bea 
    H_g=\int du d\Omega g(\Omega)\dot{A}^A\epsilon_{BA} A^B=\mathcal{O}_g.
    \eea 
    This duality flux generates the generalized EM duality transformations, i.e.,
    \begin{align}
        \delta_g A_A=i[\mathcal{O}_g,A_A].
    \end{align}

Now we can define our terminology about various transformations which extends the one in the vector theory. For geometric transformations, we have the following four kinds 
\bea  
\text{Special supertranslation (SST)}&\Leftrightarrow&\dot{f}=0,\\
\text{General supertranslation (GST)}&\Leftrightarrow&\dot{f}\not=0,\\
\text{Special superrotation (SSR)}&\Leftrightarrow&\dot{Y}^A=0,\\
\text{General superrotation (GSR)}&\Leftrightarrow&\dot{Y}^A\not=0.
\eea
Especially, a spacetime translation is a SST when $f$ obeys the equation $2\nabla_A\nabla_Bf-\gamma_{AB}\nabla^2f=0$ whose solution is $f=a_\mu n^\mu$, with $a_\mu$ constants. Similarly, a Lorentz transformation is a SSR when $Y$ satisfies the conformal Killing equation on $S^2$
\be 
\Theta_{AB}(Y)=0,\label{Theta}
\ee where we have used the symmetric traceless tensor
$\Theta_{AB}(Y)$ defined in \eqref{Thdef}.
The equation \eqref{Theta} is solved when $Y$ is a CKV.  It is clear that the transformations combining GSTs and SSRs are just Carrollian diffeomorphisms.

 Moreover, we have three kinds of duality transformations for electromagnetic theory and the gravitational theory (defined in the next section)
\bea
\text{Duality transformation (DT)}&\Leftrightarrow&g= \text{const.},\\
\text{Special super-duality transformation (SSDT)}&\Leftrightarrow&\dot g=0,\\
\text{General super-duality transformation (GSDT)}&\Leftrightarrow&\dot g\not=0.
\eea
It is worth noting that there are non-local terms when considering the variations of the vector field under GSDTs. This is similar to the case of GSRs. Therefore, we considered the algebra generated by GSTs, SSRs and  SSDTs in the vector theory \cite{Liu:2023qtr}. The same is true for the gravitational theory, as we will show in the next section.

 \section{Linearized gravity}\label{lineartheory}
 In Einstein gravity, there is no suitable definition of local stress tensor. Therefore, we will first work in the linearized gravity and regard the gravitational theory as a spin 2 tensor field theory in Minkowski background.
 \subsection{Fluxes}
   We may expand the metric around the Minkowski spacetime 
\bea 
g_{\mu\nu}=\eta_{\mu\nu}+h_{\mu\nu}.\label{gmn}
\eea Then Einstein-Hilbert action becomes the  Pauli-Fierz (PF) action 
\bea 
S_{\text{PF}}=-\frac{1}{64\pi G}\int d^4x[\partial_\mu h_{\alpha\beta}\partial^\mu h^{\alpha\beta}-\partial_\mu h\partial^\mu h+2\partial_\mu h^{\mu\nu}\partial_\nu h-2\partial_\mu h^{\mu\nu}\partial_\rho h^{\rho}_{\ \nu}].
\eea The PF action may be written as 
\bea 
\hspace{-20pt} S_{\text{PF}}=-\frac{1}{64\pi G}\int d^4 x L^{\mu_1\mu_2\cdots\mu_6}\partial_{\mu_1}h_{\mu_2\mu_3}\partial_{\mu_4}h_{\mu_5\mu_6},
\eea where the tensor $L^{\mu_1\mu_2\cdots\mu_6}$ is
\begin{align}
    L^{\mu_1\mu_2\cdots\mu_6}=&\frac{1}{2}(\eta^{\mu_1\mu_4}\eta^{\mu_2\mu_5}\eta^{\mu_3\mu_6}+\eta^{\mu_1\mu_4}\eta^{\mu_2\mu_6}\eta^{\mu_3\mu_5})-\eta^{\mu_1\mu_4}\eta^{\mu_2\mu_3}\eta^{\mu_5\mu_6}\nn\\
&+\frac{1}{2}(\eta^{\mu_1\mu_2}\eta^{\mu_3\mu_4}\eta^{\mu_5\mu_6}+\eta^{\mu_1\mu_3}\eta^{\mu_2\mu_4}\eta^{\mu_5\mu_6}+\eta^{\mu_4\mu_5}\eta^{\mu_6\mu_1}\eta^{\mu_2\mu_3}+\eta^{\mu_4\mu_6}\eta^{\mu_5\mu_1}\eta^{\mu_2\mu_3})\nn\\
&-\frac{1}{2}(\eta^{\mu_1\mu_2}\eta^{\mu_3\mu_6}\eta^{\mu_4\mu_5}+\eta^{\mu_1\mu_3}\eta^{\mu_2\mu_6}\eta^{\mu_4\mu_5}+\eta^{\mu_1\mu_2}\eta^{\mu_3\mu_5}\eta^{\mu_4\mu_6}+\eta^{\mu_1\mu_3}\eta^{\mu_2\mu_5}\eta^{\mu_4\mu_6}).
\end{align}
Properties of this tensor can be found in Appendix \ref{Lpro}. 
The action is invariant under the linearized coordinate transformation 
\be 
    h_{\mu\nu}\to h_{\mu\nu}+\partial_{\mu}\xi_{\nu}+\partial_\nu\xi_\mu.
\ee  

There are various ways to obtain the stress tensor for the linearized theory. We will accommodate the Landau-Lifshitz pseudotensor \cite{1963ZaMM...43R.287M}
\begin{align}
    (-g)T_{\text{LL}}^{\mu\nu}=\frac{1}{16\pi G}&[\partial_\lambda \mathfrak{g}^{\mu\nu}\partial_\kappa \mathfrak{g}^{\kappa\lambda}-\partial_\lambda \mathfrak{g}^{\mu\lambda}\partial_\kappa \mathfrak{g}^{\nu\kappa}+\frac{1}{2}{g}^{\mu\nu}g_{\lambda\kappa}\partial_\sigma \mathfrak{g}^{\lambda\tau}\partial_\tau\mathfrak{g}^{\kappa\sigma}\nn\\
    &-g^{\mu\lambda}g_{\kappa\tau}\partial_\sigma\mathfrak{g}^{\nu\tau}\partial_\lambda\mathfrak{g}^{\kappa\sigma}-g^{\nu\lambda}g_{\kappa\tau}\partial_\sigma\mathfrak{g}^{\mu\tau}\partial_\lambda\mathfrak{g}^{\kappa\sigma}+g_{\lambda\kappa}g^{\tau\sigma}\partial_\tau \mathfrak{g}^{\mu\lambda}\partial_\sigma\mathfrak{g}^{\nu\kappa}\nn\\
    &+\frac{1}{8}(2g^{\mu\lambda}g^{\nu\sigma}-g^{\mu\nu}g^{\lambda\sigma})(2g_{\kappa\tau}g_{\zeta\delta}-g_{\tau\zeta}g_{\kappa\delta})\partial_\lambda\mathfrak{g}^{\kappa\delta}\partial_\sigma\mathfrak{g}^{\tau\zeta} ]
\end{align}
which is widely  used in Post-Newtonian theory of gravitational theory \cite{2014grav.book.....P}. In this framework, the main variable is the so-called ``gothic'' inverse metric
\be 
\mathfrak{g}^{\mu\nu}=\sqrt{-g}g^{\mu\nu}.
\ee From the metric expansion \eqref{gmn} in linearized theory, the ``gothic'' inverse metric is 
\bea 
\mathfrak{g}^{\mu\nu}= \eta^{\mu\nu}-h^{\mu\nu}+\frac{1}{2}\eta^{\mu\nu}h
\eea up to the quadratic order of $h_{\mu\nu}$. The indices are raised using Minkowski spacetime metric $\eta^{\mu\nu}$ and $h$ is the trace of the tensor $h_{\mu\nu}$
\be 
h=h_{\mu\nu}\eta^{\mu\nu}.
\ee 
Therefore, the stress tensor corresponds to the PF action is  
\bea 
T^{\mu\nu}=\frac{1}{32\pi G}P^{\mu\nu\mu_1\cdots\mu_6}\partial_{\mu_1}h_{\mu_2\mu_3}\partial_{\mu_4}h_{\mu_5\mu_6},
\eea with the rank 8 tensor $P^{\mu\nu\mu_1\cdots\mu_6}$ defined as
\bea 
P^{\mu\nu\mu_1\cdots\mu_6}&=&2\eta^{\mu_2\mu}\eta^{\mu_3\nu}\eta^{\mu_1\mu_5}\eta^{\mu_4\mu_6}-3\eta^{\mu_1\mu_4}\eta^{\mu\mu_2}\eta^{\nu\mu_3}\eta^{\mu_5\mu_6}-2\eta^{\mu_1\mu_3}\eta^{\mu_2\mu}\eta^{\mu_4\mu_6}\eta^{\mu_5\nu}\nn\\&&+2\eta^{\mu_1\mu_4}\eta^{\mu_2\mu_6}\eta^{\mu\mu_3}\eta^{\nu\mu_5}+\eta^{\mu\mu_1}\eta^{\mu_2\mu_4}\eta^{\nu\mu_3}\eta^{\mu_5\mu_6}+\eta^{\nu\mu_1}\eta^{\mu_2\mu_4}\eta^{\mu\mu_3}\eta^{\mu_5\mu_6}\nn\\&&+2\eta^{\mu_1\mu}\eta^{\mu_2\mu_3}\eta^{\mu_4\mu_5}\eta^{\mu_6\nu}+2\eta^{\mu_1\nu}\eta^{\mu_2\mu_3}\eta^{\mu_4\mu_5}\eta^{\mu_6\mu}-2\eta^{\mu_1\mu}\eta^{\mu_2\mu_5}\eta^{\mu_3\mu_4}\eta^{\mu_6\nu}-2\eta^{\mu_1\nu}\eta^{\mu_2\mu_5}\eta^{\mu_3\mu_4}\eta^{\mu_6\mu}\nn\\&&-2\eta^{\mu_1\mu}\eta^{\mu_2\mu_3}\eta^{\mu_4\nu}\eta^{\mu_5\mu_6}+\eta^{\mu_1\mu}\eta^{\mu_2\mu_5}\eta^{\mu_3\mu_6}\eta^{\mu_4\nu}-2\eta^{\mu\nu}\eta^{\mu_1\mu_5}\eta^{\mu_2\mu_3}\eta^{\mu_4\mu_6}+\frac{3}{2}\eta^{\mu\nu}\eta^{\mu_1\mu_4}\eta^{\mu_2\mu_3}\eta^{\mu_5\mu_6}\nn\\&&-\frac{1}{2}\eta^{\mu\nu}\eta^{\mu_1\mu_4}\eta^{\mu_2\mu_5}\eta^{\mu_3\mu_6}+\eta^{\mu\nu}\eta^{\mu_1\mu_5}\eta^{\mu_2\mu_4}\eta^{\mu_3\mu_6}.
\eea 

In Cartesian coordinates, the gravitational field $h_{\mu\nu}$ may have the following fall-off behaviour 
\bea 
h_{\mu\nu}=\frac{H_{\mu\nu}}{r}+\frac{H_{\mu\nu}^{(2)}}{r^2}+\mathcal{O}\left(\frac{1}{r^3}\right),\quad \mu,\nu=0,1,2,3.\label{metricfalloffCart}
\eea 

Due to the diffeomorphism invariance of Einstein theory, we could choose Bondi gauge in this work. The first few orders of the metric are \cite{Barnich:2010eb}  
\bea 
h_{uu}&=&\frac{2GM}{r}+\frac{X}{r^2}+\mathcal{O}\left(\frac{1}{r^3}\right),\label{falloffhuu}\\
h_{ur}&=&\frac{\widetilde{X}}{r^2}+\mathcal{O}\left(\frac{1}{r^3}\right),\\ 
h_{uA}&=&\frac{1}{2}\nabla^B C_{AB}+\frac{1}{r}J_A+\mathcal{O}\left(\frac{1}{r^2}\right),\\
h_{AB}&=&rC_{AB}+Z_{AB}+\mathcal{O}\left(\frac{1}{r}\right)\label{falloffhAB}
\eea where $M$ is the Bondi mass aspect and $J_A$ is related to the angular momentum aspect. In those expansions, we have introduced the fields $X,\widetilde{X},Z_{AB}$ whose explicit forms are not important in this work, though we can write out 
\bea 
\widetilde{X}=\frac{C_{AB}C^{AB}}{16},\qquad Z_{AB}=\frac{1}{4}\gamma_{AB}C_{CD}C^{CD}.
\eea 
The symmetric and traceless tensor $C_{AB}$
\be 
C_{AB}=C_{BA},\qquad \gamma^{AB}C_{AB}=0
\ee 
is called shear tensor whose time derivative is referred to as news tensor 
\be 
N_{AB}=\dot{C}_{AB}=\frac{d}{du}C_{AB}.
\ee  
All the quantities $M,X, C_{AB}, J_A$ are fields defined at $\mathcal{I}^+$. As we will show later, we do not need the explicit form of $X$ and $J_A$ in this work. 

To transform the components of the metric in Bondi coordinates to Cartesian coordinates, we use the transformation law of the metric
\bea 
g_{\mu\nu}=\frac{\partial x'^\alpha}{\partial x^\mu}\frac{\partial x'^\beta}{\partial x^\nu}g'_{\alpha\beta}, \label{trans}
\eea where $\alpha,\beta=u,r,\theta,\phi$ are indices for the retarded frame and $\mu,\nu=0,1,2,3$ denote the Cartesian coordinates. We may relate the Cartesian coordinates to the retarded coordinates by 
\be 
x^\mu=\frac{1}{2}(n^\mu-\bar{n}^\mu)(u+r)+\frac{1}{2}(n^\mu+\bar{n}^\mu)r=\frac{u}{2}(n^\mu-\bar{n}^\mu)+r n^\mu.
\ee 
The partial derivatives of retarded coordinates are 
\bea 
\partial_\mu u=-n_\mu,\quad \partial_\mu r=\frac{1}{2}(n_\mu+\bar{n}_\mu),\quad \partial_\mu \theta^A=-\frac{1}{r}Y_\mu^A
\eea  where 
\bea 
Y^A_\mu=-\nabla^A n_\mu,\quad \mu=0,1,2,3.
\eea 
As mentioned above, $Y^A_i$ is the strictly conformal Killing vector on the unit sphere, and the vector $Y^A_0$ vanishes. We may use the vector $n^\mu$ and $Y^\nu_A$ to construct the antisymmetric tensor 
\be 
Y^A_{\mu\nu}=Y_\mu^A n_\nu-Y_\nu^An_\mu.
\ee This antisymmetric tensor corresponds to the six conformal Killing vectors on the unit sphere. Some properties of the vectors $n^\mu,\bar n^\mu, Y_\mu^A$ and the antisymmetric tensor $Y_{\mu\nu}^A$
are collected in Appendix \ref{ckvmink}.
The transformation law \eqref{trans} can be written down explicitly in Bondi gauge
\bea 
\hspace{-15pt}g_{\mu\nu}&=&n_\mu n_\nu(g_{uu}-g_{ur})-\frac{1}{2}(n_\mu\bar{n}_\nu+n_\nu\bar{n}_\mu)g_{ur}\nn\\
&&+\frac{1}{r}(n_\mu Y_\nu^A+n_\nu Y^A_\mu)g_{uA}+\frac{1}{r^2}Y_\mu^A Y_\nu^B g_{AB}.
\eea 
Now we can find the leading and subleading order of $h_{\mu\nu}$ 
\bea 
H_{\mu\nu}&=&2GM n_\mu n_\nu+\frac{1}{2}(n_\mu Y_\nu^A +n_\nu Y_\mu^A)\nabla^BC_{AB}+Y_\mu^AY_\nu^B C_{AB},\\
H_{\mu\nu}^{(2)}&=&\left(X-\widetilde{X}\right)n_\mu n_\nu-\widetilde{X}\eta_{\mu\nu}+(Z_{AB}+\widetilde{X}\gamma_{AB})Y^A_\mu Y^B_\nu +(n_\mu Y_\nu^A+n_\nu Y^A_\mu)J_A.
\eea Interestingly, $H_{\mu\nu}$ is orthogonal to the null vector $n^\mu$ and traceless 
\bea H_{\mu\nu}n^\nu=0,\qquad H_{\mu\nu}\eta^{\mu\nu}=0.\eea Some useful properties for these two tensors are collected in Appendix \ref{hh2}.
Using the chain rule, we find
\bea 
\partial_\rho\equiv \frac{\partial}{\partial x^\rho}=-n_\rho \partial_u+\frac{1}{2}(n_\rho+\bar{n}_\rho)\partial_r-\frac{1}{r}Y_\rho^A\partial_A,\quad \rho=0,1,2,3.
\eea Then the partial derivative of the field $h_{\nu\rho}$ is 
 \bea 
\partial_\mu h_{\nu\rho}&=&-\frac{n_\mu \dot{H}_{\nu\rho}}{r}-\frac{n_\mu\dot{H}_{\nu\rho}^{(2)}+\frac{1}{2}(n_\mu+\bar{n}_\mu)H_{\nu\rho}+Y_\mu^A\nabla_A H_{\nu\rho}}{r^2}+\mathcal{O}(r^{-3}).
\eea

The asymptotic expansion of the stress tensor is  
\bea 
T^{\mu\nu}&=&\frac{1}{32\pi G}[\frac{t^{\mu\nu}_{(2)}}{r^2}+\frac{t^{\mu\nu}_{(3)}}{r^3}+\cdots],\label{expansion}
\eea where 
\bea 
t^{\mu\nu}_{(2)}&=&P^{\mu\nu\mu_1\cdots\mu_6}{n}_{\mu_1}{n}_{\mu_4}\dot{H}_{\mu_2\mu_3}\dot{H}_{\mu_5\mu_6},\\
t^{\mu\nu}_{(3)}
&=&{P}^{\mu\nu\mu_1\cdots\mu_6}S_{\mu_1\cdots\mu_6}.
\eea Here, we have defined a rank 6 tensor
\bea 
S_{\mu_1\cdots\mu_6}&=&n_{\mu_1}n_{\mu_4}(\dot{H}_{\mu_2\mu_3}\dot{H}^{(2)}_{\mu_5\mu_6}+\dot{H}_{\mu_5\mu_6}\dot{H}^{(2)}_{\mu_2\mu_3})+\frac{1}{2}n_{\mu_1}(n_{\mu_4}+\bar{n}_{\mu_4})\dot{H}_{\mu_2\mu_3}H_{\mu_5\mu_6}\\&&+\frac{1}{2}n_{\mu_4}(n_{\mu_1}+\bar{n}_{\mu_1})\dot{H}_{\mu_5\mu_6}H_{\mu_2\mu_3}+n_{\mu_1}Y_{\mu_4}^A \dot{H}_{\mu_2\mu_3}\nabla_AH_{\mu_5\mu_6}+n_{\mu_4}Y_{\mu_1}^A \dot{H}_{\mu_5\mu_6}\nabla_AH_{\mu_2\mu_3}.\nn
\eea
After lengthy calculation, we find 
\bea 
t_{(2)}^{\mu\nu}&=&n^\mu n^\nu \dot{C}_{AB}\dot{C}^{AB},\\
t_{(3)}^{\mu\nu}&=&(n^\mu n^\nu+\frac{1}{2}(n_\mu\bar{n}_\nu+n_\nu\bar{n}_\mu)+\eta^{\mu\nu})\dot{H}^{\alpha\beta}H_{\alpha\beta}+(n^\mu Y^\nu_A+n^\nu Y^\mu_A)\dot{H}^{\alpha\beta}\nabla^A H_{\alpha\beta}\nn\\&&-2(n^\mu Y^A_\alpha\dot{H}^{\alpha\beta}\nabla_AH_\beta^{\ \nu}+n^\nu Y^A_\alpha\dot{H}^{\alpha\beta}\nabla_AH_\beta^{\ \mu})\nn\\&&+n^\mu\bar{n}^\alpha (H_{\alpha\beta}\dot{H}^{\beta\nu}-\dot{H}_{\alpha\beta}H^{\beta\nu})+n^\nu\bar{n}^{\alpha}(H_{\alpha\beta}\dot{H}^{\beta\mu}-\dot{H}_{\alpha\beta}H^{\beta\mu}).
\eea 

Now it is time to calculate the fluxes for linearized gravity theory. The energy and momentum charges radiated to $\mathcal{I}^+$ are
\bea 
Q_{\bm\xi_c}&=&\lim{}\hspace{-0.8mm}_+ \frac{1}{2}\int_{\mathcal{H}_r}r^2 du d\Omega (n^\mu+\bar{n}^\mu)T_{\mu\nu}\xi_c^\nu=-\frac{c_\mu}{32\pi G}\int  du d\Omega \frac{1}{2}(n_\nu+\bar{n}_\nu) t_{(2)}^{\mu\nu}\nn\\&=&-\frac{c_\mu}{32\pi G}\int  du  d\Omega n^\mu \dot{C}_{AB}\dot{C}^{AB}.
\eea 

The angular momentum and center-of-mass charges radiated to $\mathcal{I}^+$ are 
\bea 
Q_{\bm\xi_\omega}&=&\lim{}\hspace{-0.8mm}_+ \frac{1}{2}\int_{\mathcal{H}_r}r^2 du d\Omega (n^\mu+\bar{n}^\mu)T_{\mu\nu}\xi_\omega^\nu\nn\\&=&
-\frac{\omega_{\mu\nu}}{32\pi G}\int du d\Omega \frac{u}{2}\nabla_CY^{\mu\nu C}\dot{C}_{AB}\dot{C}^{AB}+\frac{\omega_{\mu\nu}}{32\pi G}\int du d\Omega Y^{\mu\nu A} H_A(C,\dot{C})
\eea 
where the co-vector $H_A(C,\dot C)$ with explicit expression
\bea 
H_A(C,\dot{C})=\frac{1}{2}(C_{BC}\nabla_A\dot{C}^{BC}-\dot{C}^{BC}\nabla_AC_{BC})+\nabla_B(\dot{C}^{BC}C_{AC}-C^{BC}\dot{C}_{AC}),
\eea is the hard Lorentz operator \cite{Campiglia:2015qka}. We have used  the identities in Appendix \ref{hh2} 
in the derivation. In addition, we have discarded the total  derivative terms 
\be 
Y^{\mu\nu A}\frac{d}{du}(C^{BC}\nabla_B C_{AC}-\frac{1}{2}C^{BC}\nabla_AC_{BC})
\ee through integration by parts.

We now construct two flux density operators from the fluxes 
\bea 
T(u,\Omega)&=&\frac{1}{32\pi G}\dot{C}_{AB}\dot{C}^{AB},\label{fluxdensities1}\\ 
M_{A}(u,\Omega)&=&-\frac{1}{32\pi G}H_{A}(C,\dot{C})\label{fluxdensities2}
\eea 
similar to the scalar and vector theories. 
We use the flux density operators $T(u,\Omega)$ and $M_A(u,\Omega)$ to construct the following smeared operators
\bea 
\mathcal{T}_f&=&\int du d\Omega f(u,\Omega)T(u,\Omega),\label{smTf}\\ 
\mathcal{M}_Y&=&\int du d\Omega Y^A(u,\Omega)M_A(u,\Omega).\label{smMY}
\eea 

 \subsection{Quantization}\label{sa}
 The smeared operators \eqref{smTf} and \eqref{smMY} are defined on the Carrollian manifold $\mathcal{I}^+$. The symplectic form \cite{Lee:1990nz, Wald:1999wa} at a hypersurface $\mathcal{H}$ from the bulk theory  is 
 \bea 
\bm\Omega^{\mathcal{H}}(\delta g;\delta g)=\frac{1}{16\pi G}\int_{\mathcal{H}} (d^3x)_\mu Q^{\mu\nu\rho\sigma\lambda\kappa}\delta g_{\nu\rho}\wedge\nabla_{\sigma}\delta g_{\lambda\kappa}
\eea with 
\bea 
Q^{\mu\nu\rho\sigma\lambda\kappa}=g^{\mu\lambda}g^{\nu\kappa}g^{\rho\sigma}-\frac{1}{2}g^{\mu\sigma}g^{\nu\lambda}g^{\rho\kappa}-\frac{1}{2}g^{\mu\nu}g^{\rho\sigma}g^{\lambda\kappa}-\frac{1}{2}g^{\mu\lambda}g^{\nu\rho}g^{\sigma\kappa}+\frac{1}{2}g^{\mu\sigma}g^{\nu\rho}g^{\lambda\kappa}.
\eea For the expansion \eqref{metricfalloffCart} we find the finite symplectic form at $\mathcal{I}^+$ 
 \bea 
 \bm\Omega(\delta_1 C,\delta_2C;C)=\lim{}\hspace{-0.8mm}_+ \bm\Omega^{\mathcal{H}_r}(\delta g,\delta g;g)=\frac{1}{32\pi G}\int du {d\Omega}\  \delta C_{AB}\wedge \delta\dot{C}^{AB}.\label{sympformgra} 
 \eea It follows that the standard commutators are \cite{Ashtekar:1981bq, Ashtekar:1981sf, Ashtekar:1987tt} 
\bea 
\ [C_{AB}(u,\Omega),C_{CD}(u',\Omega')]&=&8\pi G i P_{ACDB}\alpha(u-u')\delta(\Omega-\Omega') ,\label{cc}\\
\ [C_{AB}(u,\Omega),\dot{C}_{CD}(u',\Omega')]&=&8\pi G i P_{ACDB}\delta(u-u')\delta(\Omega-\Omega'),\label{ccd}\\
\ [\dot{C}_{AB}(u,\Omega),\dot{C}_{CD}(u',\Omega')]&=&8\pi G i P_{ACDB}\delta'(u-u')\delta(\Omega-\Omega')\label{cdcd}
\eea 
where the rank 4 tensor $P_{ABCD}$ have been defined in the vector theory \cite{Liu:2023qtr}
\be 
P_{ABCD}=\gamma_{AB}\gamma_{CD}+\gamma_{AC}\gamma_{BD}-\gamma_{AD}\gamma_{BC}.
\ee The time dependent function $\alpha(u-u')$ has also appeared in the scalar and vector theory, and its definition was given previously in \eqref{alpha}, i.e.
\be \alpha(u-u')=\frac{1}{2}[\theta(u'-u)-\theta(u-u')].
\ee 
In the Appendix \ref{canoquant}, we obtain the same commutators \eqref{cc}-\eqref{cdcd} using mode expansion of the quantized field.
In the free vacuum, the corresponding correlators are
\bea  
\langle 0|C_{AB}(u,\Omega)C_{CD}(u',\Omega')|0\rangle&=&16\pi G P_{ACDB}\beta(u-u')\delta(\Omega-\Omega'),\\
\langle 0|C_{AB}(u,\Omega)\dot C_{CD}(u',\Omega')|0\rangle&=&16\pi G P_{ACDB}\frac{1}{4\pi(u-u'-i\epsilon)}\delta(\Omega-\Omega'),\\\langle 0|\dot C_{AB}(u,\Omega) C_{CD}(u',\Omega')|0\rangle&=&-16\pi G P_{ACDB}\frac{1}{4\pi(u-u'-i\epsilon)}\delta(\Omega-\Omega'),\\
\langle 0|\dot C_{AB}(u,\Omega)\dot C_{CD}(u',\Omega')|0\rangle&=&-16\pi G P_{ACDB}\frac{1}{4\pi(u-u'-i\epsilon)^2}\delta(\Omega-\Omega').
\eea

In order to get quantum operators, we need impose normal order for flux densities \eqref{fluxdensities1} and \eqref{fluxdensities2}
\bea 
T(u,\Omega)&=&\frac{1}{32\pi G}:\dot{C}_{AB}\dot{C}^{AB}:\ ,\\ 
M_{A}(u,\Omega)&=&-\frac{1}{32\pi G}:H_{A}(C,\dot{C}):\ .
\eea
Then we could construct smeared operators with these quantized densities as in \eqref{smTf} and \eqref{smMY}.

Now it is straightforward to find the following commutators 
\bea 
\ [\mathcal{T}_f,C_{A'B'}(u',\Omega')]&=&-if(u',\Omega')\dot C_{A'B'}(u',\Omega'),\label{comTf}\\
\ [\mathcal{M}_Y,C_{A'B'}(u',\Omega')]&=&-i\Delta_{A'B'}(Y;C;u',\Omega')+\frac{i}{2}\int du \alpha(u-u')\Delta_{A'B'}(\dot Y;C;u,\Omega')\label{comMY}
\eea where 
\bea 
\Delta_{EF}(Y;C;u,\Omega)=2Y^{A}\nabla^{D}C^{BC} \rho_{ABCDEF}+\nabla^{D}Y^{A}C^{BC}P_{ABCDEF}.\label{DeltaAB}
\eea The rank 6 tensor $P_{ABCDEF}$ takes the following form
\bea 
P_{ABCDEF}&=&\frac{1}{4}(\gamma_{AB}P_{CEFD}+\gamma_{AC}P_{BEFD}+\gamma_{AD}P_{BEFC}-\gamma_{AE}P_{FBCD}-\gamma_{AF}P_{EBCD}\nn\\
&&-\gamma_{BC}P_{AEFD}+\gamma_{EF}P_{ABCD}),
\eea 
and the tensor $\rho_{ABCDEF}$ is constructed from $P_{ABCDEF}$ by 
\be 
\rho_{ABCDEF}=\frac{1}{2}(P_{ABCDEF}+P_{AEFDBC})=\frac{1}{4}\gamma_{AD}P_{BEFC}.
\ee

\subsection{Supertranslation and superrotation generators}\label{sec33}
To interpret the operators $\mathcal{T}_f$ and $\mathcal{M}_Y$, we should compute the transformation of the shear tensor induced by supertranslation and superrotation. The result could be found in \cite{Barnich:2010eb}
\bea 
\delta_f C_{AB}&=&f\dot{C}_{AB}+\gamma_{AB}\nabla^C\nabla_Cf-2\nabla_A\nabla_Bf,\label{shearst}\\
\delta_Y C_{AB}&=&\frac{1}{2}u\nabla\cdot Y \dot C_{AB} +\frac{1}{2}u\nabla^C\nabla_C\nabla\cdot Y \gamma_{AB}-u\nabla_A\nabla_B\nabla\cdot Y+Y^C\nabla_CC_{AB}\nn\\&&+\nabla_AY^CC_{BC}+\nabla_BY^CC_{AC}-\frac{1}{2}C_{AB}\nabla\cdot Y.\label{shearsr}
\eea Unfortunately, the variation induced by diffeomorphism does not match with the commutators \eqref{comTf} and \eqref{comMY}. The mismatching problem has been noticed in the electromagnetic theory \cite{Liu:2023qtr}, where we have introduced the so-called covariant variation to solve this problem. The covariant variation of any  tensor field on $\mathcal{I}^+$ is denoted as
\bea 
\delta\hspace{-6pt}\slash_{f/Y} (\cdots)\eea 
The $\cdots$  in parenthesis is any well defined field on $\mathcal{I}^+$. The subscript $f$ (or $Y$) refers to supertranslation (or superrotation). 
We use a slash to distinguish it from the original variation induced by Lie derivative. 
They are called covariant due to the following conditions.
\begin{itemize}
\item Linearity. For any scalar fields $f$ and $g$, any vector fields $Y^A$ and $Z^A$ and any constants $c_1,c_2$, we have
\bea 
\delta\hspace{-6pt}\slash_{c_1 f+c_2 g}(\cdots)&=&c_1\delta\hspace{-6pt}\slash_{f}(\cdots)+c_2\delta\hspace{-6pt}\slash_{g}(\cdots),\\
\delta\hspace{-6pt}\slash_{c_1 Y+c_2 Z}(\cdots)&=&c_1\delta\hspace{-6pt}\slash_{Y}(\cdots)+c_2\delta\hspace{-6pt}\slash_{Z}(\cdots).
\eea Also, for any two fields $F_1$ and $F_2$ of the same type, the covariant variation preserves the linearity of the tensor fields
\be 
\delta\hspace{-6pt}\slash_{f/Y} (F_1+F_2)=\delta\hspace{-6pt}\slash_{f/Y} F_1+\delta\hspace{-6pt}\slash_{f/Y} F_2.
\ee 
\item Leibniz rule. For any two fields $F_1$ and $F_2$ on $\mathcal{I}^+$, their tensor product should obey the Leibniz rule
\bea  
\delta\hspace{-6pt}\slash_{f/Y}(F_1F_2)=F_2\delta\hspace{-6pt}\slash_{f/Y}F_1+F_1\delta\hspace{-6pt}\slash_{f/Y}F_2.
\eea 
\item Metric compatibility. The covariant variation of the metric should be zero
\be 
\delta\hspace{-6pt}\slash_{f/Y} \gamma_{AB}=0.\label{metriccompa}
\ee 
\item For the scalar field $\Sigma$, the variation is the variation induced by  bulk  Lie derivative
\be 
\delta\hspace{-6pt}\slash_{f/Y}\Sigma=\delta_{f/Y}\Sigma.\label{vascalar}
\ee 
\end{itemize}

In the vector theory, the supertranslation variation induced by Lie derivative  agrees with covariant variation, though the same is not true for superrotations. However, in the gravitational theory, even the variation \eqref{shearst} is not a covariant variation due to the inhomogeneous  terms without shear tensor. 
Therefore, we subtract the inhomogeneous terms and define the covariant variation of the shear tensor under supertranslation as\footnote{As a matter of fact, the inhomogeneous terms  correspond to the soft part of the BMS fluxes in the context of full Einstein gravity. We will discuss this issue in section \ref{BMScharge} where we compare our fluxes at the linear level with the ones at the full level.}
\be 
\delta\hspace{-6pt}\slash_{f}C_{AB}=\delta_f C_{AB}-\text{inhomogeneous terms}=f\dot{C}_{AB}.
\ee Then we find what we need
\be 
\ i[\mathcal{T}_f,C_{AB}]=\delta\hspace{-6pt}\slash_{f}C_{AB},
\ee and could identify the operator $\mathcal{T}_f$ with the supertranslation generators. It is worth noting that the inhomogeneous terms in \eqref{shearst} vanishes for translations, namely $f=a_\mu n^\mu$, with $a_\mu$ constants. The same is true for \eqref{shearsr} whose inhomogeneous terms vanish for Lorentz transformations, i.e. $Y^A=\omega^{\mu\nu}Y^A_{\mu\nu}$ with $\omega^{\mu\nu}$ constants.

Now we will focus on the superrotation. 
We not only should subtract the inhomogeneous terms in \eqref{shearsr}, but also need add terms from connections
\bea 
\delta\hspace{-6pt}\slash_{Y}C_{AB}=\delta_YC_{AB}-\Gamma_{A}^{\ C}(Y)C_{CB}-\Gamma_B^{\ C}(Y)C_{AC}-\text{inhomogeneous terms}.
\eea The connection $\Gamma_{AB}(Y)$ can be chosen as a symmetric tensor which has been found in \cite{Liu:2023qtr}
\be 
\Gamma_{AB}(Y)=\frac{1}{2}\Theta_{AB}(Y).
\ee Therefore, we get
\bea 
\delta\hspace{-6pt}\slash_{Y}C_{AB}=\frac{1}{2}u\nabla_CY^C\dot{C}_{AB}+Y^C\nabla_C C_{AB}+\nabla_{[C}Y_{A]}C_{B}^{\ C}+\nabla_{[C}Y_{B]}C_A^{\ C}+\frac{1}{2}C_{AB}\nabla_CY^C.
\eea Now we observe 
\bea 
\delta\hspace{-6pt}\slash_{ f=\frac{1}{2}u\nabla_CY^C}C_{AB}&=&\frac{1}{2}u\nabla_CY^C \dot{C}_{AB},\
\eea 
 and notice that \eqref{DeltaAB} can be rewritten as
\begin{align}
    \Delta_{AB}(Y;C;u,\Omega)=&Y^C\nabla_C C_{AB}+\nabla_{[C}Y_{A]}C_{B}^{\ C}+\nabla_{[C}Y_{B]}C_A^{\ C}+\frac{1}{2}C_{AB}\nabla_CY^C.\label{Deltaab}
\end{align}

Hence, we find
\bea 
\delta\hspace{-6pt}\slash_{Y}C_{AB}=i[\mathcal{M}_Y,C_{AB}] +i[\mathcal{T}_{f=\frac{1}{2}u\nabla_CY^C},C_{AB}]
\eea for $\dot Y=0$. The second term has been identified as the contribution of a general supertranslation, so we will say that $\mathcal{M}_Y$ generates superrotations.

\subsection{Commutation relations}\label{seccom}
Since $\mathcal{T}_f$ and $\mathcal{M}_Y$ are generators of supertranslation and superrotation, we may compute the following commutators\footnote{The details can be found in the Appendix \ref{commutator}}
\begin{subequations}\label{commutators}
    \bea 
\ [\mathcal{T}_{f_1},\mathcal{T}_{f_2}]&=&C_T(f_1,f_2)+i\mathcal{T}_{f_1\dot f_2-f_2\dot f_1},\label{Tf1Tf2}\\
\ [\mathcal{T}_f,\mathcal{M}_Y]&=&-i\mathcal{T}_{Y^A\nabla_A f}+i\mathcal{M}_{f\dot Y}+i\mathcal{O}_{\dot Y^A\nabla^B f \epsilon_{BA}}+\frac{i}{4}\mathcal{Q}_{\frac{d}{du}(\dot Y^A\nabla_A f)},\\
\ [\mathcal{T}_f,\mathcal{O}_g]&=&i \mathcal{O}_{f\dot g},\\
\ [\mathcal{M}_Y,\mathcal{M}_Z]&=&C_M(Y,Z)+i\mathcal{M}_{[Y,Z]}+2i\mathcal{O}_{o(Y,Z)}+N_M(\dot Y,\dot Z),\\
\ [\mathcal{M}_Y,\mathcal{O}_g]&=&C_{MO}(Y,g)+i\mathcal{O}_{Y^A\nabla_A g}+N_{MO}(\dot Y,\dot g),\\
\ [\mathcal{O}_{g_1},\mathcal{O}_{g_2}]&=&C_O(g_1,g_2)+N_O(\dot g_1,\dot g_2).\label{OgOh}
\eea
\end{subequations}
We find two new smeared operators on the right hand side of the commutators. The first operator $\mathcal{Q}_h$  is constructed from the square of the shear tensor
\bea 
\mathcal{Q}_h&=&\frac{1}{32\pi G}\int du d\Omega h(u,\Omega):C_{AB} C^{AB}:\ .
\eea Similar operator has also appeared in the scalar and vector theory. We can not find a physical interpretation for it, since its commutators with $C_{AB}$ is totally non-local. Therefore, we do not care the commutators between $\mathcal{Q}_h$ and other operators here. Another smeared operator is 
\bea 
\mathcal{O}_g&=&\frac{1}{32\pi G}\int du d\Omega g(u,\Omega):\dot{C}^{AB}C_{B}^{\ C}:\epsilon_{CA}\nn\\
&=&\frac{1}{32\pi G}\int du d\Omega  g(u,\Omega):\dot{C}_{AB}C_{CD}:Q^{ABCD}.
\eea   We have defined a rank 4 tensor 
\bea 
Q_{ABCD}=\frac{1}{4}(\gamma^{BC}\epsilon^{DA}+\gamma^{AC}\epsilon^{DB}+\gamma^{BD}\epsilon^{CA}+\gamma^{AD}\epsilon^{CB})
\eea at the second step. This operator is parity odd on the sphere and there is a similar operator in the electromagnetic theory. Its commutator with the shear tensor is 
\bea 
\ [\mathcal{O}_g,C_{A'B'}(u',\Omega')]=-i\Delta_{A'B'}(g;C;u',\Omega')+\frac{i}{2}\int du \alpha(u-u')\Delta_{A'B'}(\dot g;C;u,\Omega'),\eea  where 
\bea 
\Delta_{AB}(g;C;u,\Omega)=g(u,\Omega)Q_{ABCD}C^{CD}(u,\Omega).
\eea 
 We will discuss its physical meaning later. Here we have included its commutators with supertranslation and superrotation operators as well as itself. 
 
 There are three non-local terms appearing in \eqref{commutators} which read 
\begin{subequations}
    \bea 
N_{M}(\dot Y,\dot Z)&=&\frac{i}{64\pi G}\int du du'd\Omega \alpha(u'-u)\Delta_{AB}(\dot Y;C;u',\Omega)\Delta^{AB}(\dot Z;C;u,\Omega),\\
N_{MO}(\dot Y,\dot g)
&=&\frac{i}{64\pi G}\int du du'd\Omega\alpha(u'-u)\Delta_{GH}(\dot g;C;u,\Omega)\Delta^{GH}(\dot Y;C;u',\Omega),\\
N_{O}(\dot g_1,\dot g_2)&=&\frac{i}{64\pi G}\int du du'd\Omega\alpha(u'-u)\Delta_{CD}(\dot g_2;C;u,\Omega)\Delta^{CD}(\dot g_1;C,u',\Omega).
\eea 
\end{subequations}
Moreover, the related central terms are listed below
\begin{subequations}
    \bea 
C_T(f_1,f_2)&=&-\frac{ic}{24\pi}\mathcal{I}_{f_1\dddot {f}_{\hspace{-3.5pt}2}-f_2\dddot {f}_{\hspace{-3.5pt}1}},\\
C_M(Y,Z)&=&\int du du'd\Omega d\Omega' Y^A(u,\Omega)Z^{B'}(u',\Omega')\Lambda_{AH'}^{(2)}(\Omega-\Omega')\eta(u-u'),\\
C_{MO}(Y,g)&=&-4c\int du du'd\Omega Y^A(u,\Omega)\nabla^Bg(u',\Omega) \epsilon_{AB}\eta(u-u'),\\
C_O(g_1,g_2)&=&4c\int du du'd\Omega \eta(u-u')g_1(u,\Omega)g_2(u',\Omega),
\eea
\end{subequations}
 where 
\bea
\Lambda^{(2)}_{AH'}(\Omega,\Omega')&=&P_{ABCDEF}P_{H'I'J'K'L'M'}[P^{BI'J'C}P^{EL'M'F}\delta(\Omega-\Omega')\nabla^D\nabla^{K'}\delta(\Omega-\Omega')\nn\\
&&-P^{EI'J'F}P^{BL'M'C}\nabla^D\delta(\Omega-\Omega')\nabla^{K'}\delta(\Omega-\Omega')].
\eea 
Besides, we use $c$ to denote the Dirac function on sphere with argument equalling to zero, i.e., $c=\delta^{(2)}(0)$. The central charge $C_T(f_1,f_2)$ is exactly twice as much as the one in real scalar case, as one expects. This is due to the fact that the number of the propagating degrees of freedom for linearized gravity is 2. The same phenomenon appears in the vector theory.

 Notice that the \eqref{Tf1Tf2} is actually a higher dimensional Virasoro algebra. One can perform Fourier transformation for $f$  
\begin{align}
  f(u,\Omega)=\sum_{\ell,m}\int_{-\infty}^\infty d\omega c_{\omega,\ell,m}f_{\omega,\ell,m}
\end{align}
where $f_{\omega,\ell,m}=e^{-i\omega u}Y_{\ell,m}(\Omega)$.  The modes of \(\mathcal{T}_f\) become
\begin{align}
  \mathcal{T}_{\omega,\ell,m}=\int du d\Omega e^{-i\omega u}Y_{\ell,m}(\Omega)T(u,\Omega).\label{lf2}
\end{align}
Therefore, \eqref{Tf1Tf2} implies\footnote{We have corrected the corresponding  central term of higher dimensional Virasoro algebra in  our previous papers \cite{Liu:2022mne,Liu:2023qtr} by inserting a $(-1)^m$ and replacing $\delta_{m,m'}$ to $\delta_{m,-m'}$.}
\begin{align}
  &[\mathcal{T}_{\omega, \ell, m}, \mathcal{T}_{\omega', \ell', m'}]=(\omega'-\omega)\sum_{L=|\ell-\ell'|}^{\ell+\ell'}\sum_{M=-L}^Lc_{\ell, m; \ell', m'; L, M}\mathcal{T}_{\omega+\omega', L, M}-\frac{\omega^3}{6}(-1)^m\delta^{(2)}(0)\delta(\omega+\omega')\delta_{\ell\ell'}\delta_{m,-m'},
\end{align}
where $ c_{\ell, m; \ell', m'; L, M}$ are related to Clebsch-Gordan coefficients. They can be explicitly given by Wigner $3j$-Symbols as follows
\begin{align}
  c_{\ell_1,m_1;\ell_2,m_2;L,M}=(-1)^M\sqrt{\frac{(2\ell_1+1)(2\ell_2+1)(2L+1)}{4\pi}}
  \left(\begin{array}{ccc}
    \ell_1&\ell_2&L\\0&0&0
  \end{array}\right)
  \left(\begin{array}{ccc}
    \ell_1&\ell_2&L\\ m_1&m_2&-M
  \end{array}\right).
\end{align}

\subsubsection*{Truncations}
The above algebra is not closed due to the non-local terms. We need to eliminate them.  It is easy to find that when 
\be 
\dot Y=\dot Z=\dot g_1=\dot g_2=0,
\ee 
 all the non-local terms vanish. Moreover, three of the central terms $C_M, C_{MO},C_O$, and the physically meaningless operator $\mathcal{Q}_h$ all disappear. In this  case, we obtain a closed Lie algebra 
\begin{subequations}
    \bea 
\ [\mathcal{T}_{f_1},\mathcal{T}_{f_2}]&=&C_T(f_1,f_2)+i\mathcal{T}_{f_1\dot f_2-f_2\dot f_1},\\
\ [\mathcal{T}_f,\mathcal{M}_Y]&=&-i\mathcal{T}_{Y^A\nabla_A f},\\
\ [\mathcal{T}_f,\mathcal{O}_g]&=&0,\\
\ [\mathcal{M}_Y,\mathcal{M}_Z]&=&i\mathcal{M}_{[Y,Z]}+2i\mathcal{O}_{o(Y,Z)},\label{mymz}\\
\ [\mathcal{M}_Y,\mathcal{O}_g]&=&i\mathcal{O}_{Y^A\nabla_A g},\\
\ [\mathcal{O}_{g_1},\mathcal{O}_{g_2}]&=&0.
\eea 
\end{subequations}

This Lie algebra is our main result whose structure is almost the same as the one in the electromagnetic theory, except the factor 2 before the operator $\mathcal{O}$ in  \eqref{mymz}. After rescaling the operator $\mathcal{O}\to \frac{1}{2}\mathcal{O}$,  the algebra is isomorphic to the one in the electromagnetic theory. However, we will not try to rescale the operator, since the original commutators \eqref{Tf1Tf2}-\eqref{OgOh} are not isomorphic to the one in the electromagnetic theory  even with this rescaling. 

Actually, it is possible to write the closed Lie algebra with an arbitrary parameter $s$
\begin{subequations}
    \bea 
\ [\mathcal{T}_{f_1},\mathcal{T}_{f_2}]&=&C_T(f_1,f_2)+i\mathcal{T}_{f_1\dot f_2-f_2\dot f_1},\label{Tf1Tf2s}\\
\ [\mathcal{T}_f,\mathcal{M}_Y]&=&-i\mathcal{T}_{Y^A\nabla_A f},\\
\ [\mathcal{T}_f,\mathcal{O}_g]&=&0,\\
\ [\mathcal{M}_Y,\mathcal{M}_Z]&=&i\mathcal{M}_{[Y,Z]}+i s\mathcal{O}_{o(Y,Z)},\\
\ [\mathcal{M}_Y,\mathcal{O}_g]&=&i\mathcal{O}_{Y^A\nabla_A g},\\
\ [\mathcal{O}_{g_1},\mathcal{O}_{g_2}]&=&0.\label{OgOhs}
\eea
\end{subequations}
For $s=0,1,2$, the Lie algebra corresponds to the complex scalar, electromagnetic and gravitational theory, respectively. It may  be correct for the theory of arbitrary  spin.

 \subsection{Duality operator}
 Duality invariance of Maxwell equation leads to the introduction of magnetic monopole and the quantization of electric charge \cite{Dirac:1931mon}. It has been elaborated in  non-Abelian gauge theories by \cite{1976PhRvD..13.1592D}. In the context of linearized gravity, the duality symmetry has been discussed in  
 \cite{Henneaux:2004jw}. In de Sitter spacetime and anti-de Sitter spacetime, the duality symmetry of linearized gravity is discussed in \cite{2005JHEP...11..025J, 2019FrP.....7..188H}. Now we are going to derive the flux operator corresponding to duality transformations in the linearized gravity. 
 
 Similar to what has been done in the electromagnetic theory, we introduce the dual Riemann tensor 
 \bea 
 \widetilde{R}_{\mu\nu\rho\sigma}=-\frac{1}{2}\epsilon_{\mu\nu}^{\hspace{10pt}\alpha\beta}R_{\alpha\beta\rho\sigma},\label{dualriem}
 \eea 
 where the linearized Riemann tensor reads
 \begin{align}
     {R}_{\mu\nu\rho\sigma}=\frac{1}{2}(\partial_\rho\partial_\nu h_{\mu\sigma}-\partial_\rho\partial_\mu h_{\nu\sigma}-\partial_\sigma\partial_\nu h_{\mu\rho}+\partial_\sigma \partial_\mu h_{\nu\rho}).\label{riemann}
 \end{align}
 As a consequence of the Bianchi identity  
 \be 
 R_{\mu[\nu\rho\sigma]}=0,
 \ee the dual Ricci tensor is zero 
 \bea 
 \widetilde{R}_{\mu\nu}=0.
 \eea 
 A gravitational duality is a $SO(2)$ transformation which rotates the Riemann tensor and its dual 
 \bea 
 &&R'_{\mu\nu\rho\sigma}=\cos\varphi R_{\mu\nu\rho\sigma}+\sin\varphi \widetilde{R}_{\mu\nu\rho\sigma},\\ &&\widetilde{R}_{\mu\nu\rho\sigma}'=-\sin\varphi R_{\mu\nu\rho\sigma}+\cos\varphi \widetilde{R}_{\mu\nu\rho\sigma},
 \eea where $\varphi$ is a constant rotation angle. 
One can show that the linearized equation of motion is invariant under the dual transformation 
\bea 
&R'_{\mu\nu}=\cos\varphi R_{\mu\nu}+\sin\varphi \widetilde{R}'_{\mu\nu}=0,\\
&\widetilde{R}_{\mu\nu}'=-\sin\varphi R_{\mu\nu}+\cos\varphi\widetilde{R}_{\mu\nu}=0.
 \eea 
 
The next step is to introduce a dual gravitational field $\widetilde{h}_{\mu\nu}$ such that 
\bea
\widetilde{R}_{\mu\nu\rho\sigma}=\frac{1}{2}(\partial_\rho\partial_\nu \widetilde h_{\mu\sigma}-\partial_\rho\partial_\mu\widetilde h_{\nu\sigma}-\partial_\sigma\partial_\nu\widetilde h_{\mu\rho}+\partial_\sigma \partial_\mu\widetilde h_{\nu\rho}).
\eea Then the dual transformation may be written as a $SO(2)$ rotation between $h_{\mu\nu}$ and $\widetilde{h}_{\mu\nu}$
\bea 
h'_{\mu\nu}=\cos\varphi h_{\mu\nu}+\sin\varphi \widetilde{h}_{\mu\nu},\qquad \widetilde{h}_{\mu\nu}'=-\sin\varphi h_{\mu\nu}+\cos\varphi \widetilde{h}_{\mu\nu}.
\eea The dual Riemann tensor is invariant under the dual coordinate transformation 
\bea 
\delta_{\widetilde{\xi}}\widetilde{h}_{\mu\nu}=\partial_\mu \widetilde{\xi}_\nu+\partial_\nu\widetilde{\xi}_\mu.
\eea We may expand the dual gravitational field near $\mathcal{I}^+$ 
\bea 
\widetilde{h}_{\mu\nu}=\frac{\widetilde{H}_{\mu\nu}}{r}+\mathcal{O}\left(\frac{1}{r^2}\right),\quad \mu,\nu=0,1,2,3\label{wdthexp}
\eea where 
\bea 
\widetilde{H}_{\mu\nu}&=&2G\widetilde{M} n_\mu n_\nu+\frac{1}{2}(n_\mu Y_\nu^A +n_\nu Y_\mu^A)\nabla^B\widetilde{C}_{AB}+Y_\mu^AY_\nu^B \widetilde{C}_{AB}.
\eea We will call $\widetilde{M}$ the dual Bondi mass aspect and $\widetilde{C}_{AB}$ the dual shear tensor which have been extensively studied in the literature \cite{Godazgar:2018qpq,Godazgar:2019dkh,Godazgar:2020kqd,Oliveri:2020xls,Freidel:2021fxf,Freidel:2021qpz,PhysRevLett.129.061101}. 

Combining the expansion \eqref{wdthexp} with the duality relation \eqref{dualriem}, we find the following relation at the leading order near $\mathcal{I}^+$
\bea 
-Y_{\rho\sigma}^AY_{\mu\nu}^B\ddot{\widetilde{C}}_{AB}=\frac{1}{2}\epsilon_{\mu\nu}^{\hspace{10pt}\alpha\beta}Y_{\alpha\beta}^AY_{\rho\sigma}^B\ddot{C}_{AB}. \label{relationCtildeC}
\eea Considering the identity \eqref{2ymunu}, the relation \eqref{relationCtildeC} is satisfied by imposing the duality condition 
\bea 
\widetilde{C}_{AB}=Q_{AB}^{\hspace{13pt}CD}C_{CD}.
\eea

As a consequence of the duality invariance, we may construct the conserved current\footnote{The conserved current may be derived similar to the electromagnetic duality current. The conserved charge is the difference of the numbers of gravitons with positive and negative helicity.  One can find more details on the derivation in Appendix \ref{dualcur}.}  
\bea 
j^\mu_{\text{duality}}=\frac{1}{64\pi G}L^{\mu\rho\sigma\mu_4\mu_5\mu_6}(h_{\rho\sigma}\partial_{\mu_4}\widetilde{h}_{\mu_5\mu_6}-\widetilde{h}_{\rho\sigma}\partial_{\mu_4}h_{\mu_5\mu_6}).
\eea Using the fall-off conditions, we find the helicity  flux  leaking to $\mathcal{I}^+$ at time $u$
\bea 
\lim{}\hspace{-0.8mm}_+ \frac{1}{2}\int_{S^2} d\Omega r^2 (n_\mu+\bar{n}_\mu)j^\mu_{\text{duality}}=\frac{1}{32\pi G}\int_{S^2} d\Omega \dot{C}^{AB}Q_{ABCD}C^{CD}.
\eea We may read out the helicity flux density operator (after quantization)
\bea 
O(u,\Omega)=\frac{1}{32\pi G}:\dot{C}^{AB}Q_{ABCD}C^{CD}:\ ,
\eea from which we can define the smeared operator 
\bea 
\mathcal{O}_g=\int du d\Omega g(u,\Omega)O(u,\Omega).
\eea 
From now on, we will call $\mathcal{O}_g$ the duality operator.

 \subsection{Hamiltonians related to Carrollian diffeomorphisms and duality transformations}\label{canochargegrav}
In this subsection, we will use the boundary symplectic form \eqref{sympformgra} and the relation \eqref{vacharge} in covariant phase space formalism to compute the Hamiltonians corresponding to the Carrollian diffeomorphisms and gravitational duality transformations. The interior product in \eqref{vacharge} should be modified to \eqref{modiinter} similar to electromagnetic theory.  We emphasize that our computation of Hamiltonians is in the boundary Carrollian field theory.  

\begin{enumerate}
    \item For Carrollian diffeomorphism generated by $\bm\xi=\bm\xi_f=f(u,\Omega)\partial_u$, 
    \bea 
    \delta H_f&=&i_{\bm\xi_f}\frac{1}{32\pi G}\int du d\Omega \delta C_{AB}\wedge \delta \dot{C}^{AB}\nn\\&=&\frac{1}{32\pi G}\int du d\Omega \{f(u,\Omega)\dot{C}_{AB}\delta \dot{C}^{AB}-\delta C_{AB}\frac{d}{du}[f(u,\Omega)\dot{C}^{AB}]\}\nn\\&=&\frac{1}{16\pi G}\int du d\Omega f(u,\Omega)\dot{C}_{AB}\delta \dot C^{AB}\nn\\&=& \frac{1}{32\pi G}\delta\int du d\Omega f(u,\Omega)\dot{C}_{AB}\dot{C}^{AB}.
    \eea Therefore, we find an integrable flux
    \bea 
    H_f=\frac{1}{32\pi G}\int du d\Omega f(u,\Omega)\dot{C}_{AB}\dot{C}^{AB}\equiv\mathcal{T}_f.\label{Hfgrav}
    \eea 
    \item For Carrollian diffeomorphism generated by $\bm\xi=\bm\xi_Y=Y^A(\Omega)\partial_A$, 
    \bea 
    \delta H_{Y}-\delta H_{f=\frac{1}{2}u\nabla\cdot Y}&=&i_{\bm\xi_Y}\frac{1}{32\pi G}\int du d\Omega \delta{C}_{AB}\wedge\delta \dot{C}^{AB}\nn\\&=&\frac{1}{32\pi G}\int du d\Omega [\Delta_{AB}(Y;C;u,\Omega)\delta\dot{C}^{AB}-\delta C_{AB}\Delta^{AB}(Y;\dot C;u,\Omega)]\nn\\&=&\frac{1}{32\pi G}\delta \int du d\Omega \dot{C}^{AB}\Delta_{AB}(Y;C;u,\Omega).\label{integrablechargeY}
    \eea The integrable flux reads out as 
    \bea 
    H_Y-H_{f=\frac{1}{2}u\nabla\cdot Y}=\frac{1}{32\pi G}\int du d\Omega \dot{C}^{AB}\Delta_{AB}(Y;C;u,\Omega)\equiv\mathcal{M}_Y.\label{HYgra}
    \eea When $Y^A$ is time dependent, there is an additional term 
    \bea 
    \delta H_Y-\delta H_{f=\frac{1}{2}u\nabla\cdot Y}=(\cdots)-\frac{1}{32\pi G}\int du d\Omega \delta {C}_{AB}\Delta^{AB}(\dot Y; C;u,\Omega).
    \eea In this expression, the $(\cdots)$ denotes the integrable flux \eqref{integrablechargeY}. The additional term is not integrable. 
    \item For the gravitational SSDT generated by a smooth function $g(\Omega)$ on $S^2$, 
    \bea 
    \delta_g C_{AB}=g(\Omega)Q_{ABCD}C^{CD}=\Delta_{AB}(g;C;u,\Omega),
    \eea we find 
    \bea 
    \delta H_g&=&\frac{1}{32\pi G}\int du d\Omega [\Delta_{AB}(g;C;u,\Omega)\delta\dot{C}^{AB}-\delta C_{AB}\Delta^{AB}(g;\dot C;u,\Omega)]\nn\\&=&\frac{1}{32\pi G} \delta \int du d\Omega \dot{C}^{AB}\Delta_{AB}(g;C;u,\Omega).
    \eea The integrable flux is 
    \bea 
    H_g=\frac{1}{32\pi G}\int du d\Omega \dot{C}^{AB}\Delta_{AB}(g;C;u,\Omega)\equiv\mathcal{O}_{g}.\label{Hggra}
    \eea When $g$ is time dependent, the corresponding variation $\delta H_g$ is not integrable. 
\end{enumerate}
We have shown that the corresponding Hamiltonians for Carrollian diffeomorphisms and gravitational duality transformations match with the ones derived from radiation fluxes. More interestingly, the following formula
\bea 
H_{\zeta}=\int du d\Omega \dot{C}^{AB}\delta\hspace{-6pt}\slash_\zeta C_{AB}\label{Hzeta}
\eea are exactly \eqref{Hfgrav}, \eqref{HYgra} and \eqref{Hggra} for $\zeta=f(u,\Omega),Y^A(\Omega)$ and $g(\Omega)$, respectively. The formula may be extended to 
\bea 
H_{\zeta}=\int du d\Omega \dot{F}\delta\hspace{-6pt}\slash_\zeta F
\eea 
for general Carrollian field theories.

\section{Relation to BMS  fluxes in nonlinear Einstein gravity}\label{BMScharge}
The previous fluxes were first derived at the linear level from the bulk Landau-Lifshitz pseudotensor. The energy and momentum fluxes are quadratic in news tensor, while the angular momentum and center-of-mass fluxes happen to be the hard Lorentz operator after discarding total time derivatives, which all take the same form as fluxes in the scalar and vector theory \cite{Liu:2022mne,Liu:2023qtr}. Moreover, we use the formula \eqref{vacharge} to re-derive the aforementioned fluxes, or called Hamiltonians in the sense of generating boundary transformations. In this section, we will discuss the relation between our flux operators and the BMS fluxes defined in the context of Einstein gravity \cite{Barnich:2011mi,Hawking:2016sgy,Flanagan:2015pxa,Freidel:2021fxf,Compere:2018ylh,Compere:2019bua,2020JHEP...10..116C,Compere:2020lrt,Ruzziconi:2020cjt,Barnich:2021dta,Donnay:2021wrk,Fiorucci:2021pha,Donnay:2022hkf}. 

\subsection{Backgrounds}
An asymptotically flat spacetime in general relativity is a solution of Einstein equation with an external source 
\bea 
G_{\mu\nu}=8\pi G T_{\mu\nu}
\eea where $G_{\mu\nu}$ is the Einstein tensor and $T_{\mu\nu}$ is the stress tensor for matters. 
The stress tensor is assumed to satisfy the fall-off conditions near $\mathcal{I}^+$
\bea 
T_{uu}&=&\frac{t_{uu}(u,\Omega)}{r^2}+\mathcal{O}\left(\frac{1}{r^3}\right),\\
T_{ur}&=&\frac{t_{ur}(u,\Omega)}{r^4}+\mathcal{O}\left(\frac{1}{r^5}\right),\\
T_{uA}&=&\frac{t_{uA}(u,\Omega)}{r^2}+\mathcal{O}\left(\frac{1}{r^3}\right),\\
T_{rr}&=&\frac{t_{rr}(u,\Omega)}{r^4}+\mathcal{O}\left(\frac{1}{r^5}\right),\\
T_{rA}&=&\frac{t_{rA}(u,\Omega)}{r^3}+\mathcal{O}\left(\frac{1}{r^4}\right),\\
T_{AB}&=&\frac{t_{AB}(u,\Omega)}{r}+\mathcal{O}\left(\frac{1}{r^2}\right).
\eea 
An asymptotically flat metric 
may be written in Bondi gauge as 
\bea 
ds^2=ds_{\rm M}^2+\delta g_{\mu\nu}dx^\mu dx^\nu
\eea  with $ds_{\rm M}^2$ denoting line element for Minkowski spacetime, and the components of $\delta g_{\mu\nu}$ reading
\bea \delta g_{uu}&=&\frac{2GM}{r}+\mathcal{O}\left(\frac{1}{r^2}\right),\\
\delta g_{ur}&=&\left(\frac{1}{16}C_{AB}C^{AB}+2\pi G t_{rr}\right)\frac{1}{r^2}+\mathcal{O}\left(\frac{1}{r^3}\right),\\
\delta g_{uA}&=&\frac{1}{2}\nabla^BC_{AB}+\frac{1}{2r}\left[\frac{4}{3}\bar{N}_A-\frac{1}{8}\nabla_A(C_{BC}C^{BC})\right]+\mathcal{O}\left(\frac{1}{r^2}\right),\\
\delta g_{AB}&=&rC_{AB}+\left(\frac{1}{4}\gamma_{AB}C_{CD}C^{CD}+\mathcal{D}_{AB}\right)+\mathcal{O}\left(\frac{1}{r}\right).
\eea The symmetric traceless tensor $\mathcal{D}_{AB}$ is conserved
\be 
\dot{\mathcal{D}}_{AB}=0.
\ee 
The conservation of the stress tensor $\nabla^\mu T_{\mu\nu}=0$ implies 
\bea 
\dot t_{rA}=\frac{1}{2}\nabla_A (\gamma^{BC}t_{BC}),\qquad \dot{t}_{rr}=-\gamma^{AB}t_{AB}.
\eea The Bondi mass aspect $M(u,\Omega)$ and Bondi angular momentum aspect $\bar{N}_A$ are constrained by the following equations 
\bea 
\dot{M}&=&-4\pi t_{uu}-\frac{1}{8G}\dot{C}_{AB}\dot{C}^{AB}+\frac{1}{4G}\nabla_A\nabla_B \dot{C}^{AB},\label{bondimass}\\
\dot{\bar{N}}_A&=&-8\pi (t_{uA}+\frac{1}{8}\nabla_A (\gamma^{BC}t_{BC}))+\nabla_AM+\frac{1}{4G}\nabla^B(\nabla_A\nabla^CC_{BC}-\nabla_B\nabla^C C_{AC})\nn\\&&+\frac{1}{4G}\nabla_B(\dot{C}^{BC}C_{AC})+\frac{1}{2G}\nabla_B\dot{C}^{BC} C_{AC}.\label{bondiangular}
\eea
The Bondi mass aspect and the Bondi angular momentum aspect may be used to define the BMS charges \cite{2020JHEP...10..116C}
\bea 
\mathcal{P}_f&=&\frac{1}{4\pi}\int_{S^2} d\Omega f(\Omega) M,\label{stcharge1}\\ 
\mathcal{R}_Y&=&\frac{1}{8\pi}\int_{S^2} d\Omega Y^A(\Omega)N_A,\label{srcharge2}
\eea where  $N_A$ is related to $\bar{N}_A$ by 
\bea 
N_A&=&\bar{N}_A-u\nabla_AM-\frac{1}{4G}C_{AB}\nabla_CC^{BC}-\frac{1}{16G}\nabla_A(C_{BC}C^{BC})\nn\\
&&+\frac{u}{4G}\Box \nabla^CC_{AC}-\frac{u}{4G}\nabla_B\nabla_A\nabla_C C^{BC}.\label{shiftN}
\eea We have changed the notation $\mathcal{T}$ to $\mathcal{P}$ to denote the supertranslation charge. The surface charge $\mathcal{P}_f$ associated with $f(\Omega)$ is called supermomenta. At the same time, we have written the superrotation  charge as $\mathcal{R}_Y$\footnote{In \cite{2020JHEP...10..116C}, the BMS charge with $Y^A=\gamma^{AB}\partial_B\Psi$ is called superboost charge. We will only use the concept of superrotation, without distinguishing superrotation and superboost charges.}. The Bondi mass aspect is time dependent, we may use \eqref{bondimass} to find the supermomenta 
\bea 
\mathcal{P}_f(u_i,u_f)=\frac{1}{4\pi}\int_{u_i}^{u_f} du \int_{S^2}d\Omega f(\Omega) \dot{M}(u,\Omega)
\eea where we have chosen an initial time $u_i$ and a final time $u_f$. This is the supermomenta radiated to $\mathcal{I}^+$ during the time duration $u_i<u<u_f$. The supermomentum  may be rewritten as an integral on the Carrollian manifold $\mathcal{I}^+$
\bea 
\mathcal{P}_f(u_i,u_f)=\frac{1}{4\pi}\int du d\Omega \theta(u_f-u)\theta(u-u_i)f(\Omega)\dot{M}(u,\Omega).\label{stcharge}
\eea This may be generalized slightly to the following smeared operator 
\bea 
\bar{\mathcal{T}}_f=-\frac{1}{4\pi}\int du d\Omega f(u,\Omega)\dot{M}(u,\Omega), \label{gstcharge}
\eea where we have extended the function $\theta(u_f-u)\theta(u-u_i)f(\Omega)$ to any smooth function $f(u,\Omega)$ on $\mathcal{I}^+$ which depends on the retarded time $u$. We have changed the notation from $\mathcal{P}_f$ to $\bar{\mathcal{T}}_f$ to compare with our definition of flux operators $\mathcal{T}_f$. We add a minus sign in the definition since $\dot M$ is non-positive due to the radiation. Moreover, we add a bar in $\bar{\mathcal{T}}_f$ since the quantity $\dot{M}$ in the BMS charge is from  Einstein equation. It is not necessary to have the same form as the energy flux density operator $T(u,\Omega)$ defined in the previous sections. 

Similarly, the superrotation charge \eqref{srcharge2} may be written explicitly as 
\bea 
\mathcal{R}_Y(u_i,u_f)=\frac{1}{8\pi}\int_{-\infty}^{\infty} du \int_{S^2}d\Omega\ \theta(u_f-u)\theta(u-u_i)Y^A(\Omega)\dot{N}_A(u,\Omega)
\eea and generalized to  
\bea 
\bar{\mathcal{M}}_Y=-\frac{1}{8\pi}\int du d\Omega Y^A(u,\Omega)\dot N_A(u,\Omega)-\bar{\mathcal{T}}_{f=\frac{1}{2}u\nabla_A Y^A}. \label{gsrcharge}
\eea  We have also changed the notation from $\mathcal{R}_Y$ to $\bar{\mathcal{M}}_Y$ to compare with our definition of flux operators $\mathcal{M}_Y$. The operator $\bar{\mathcal{T}}_{f=\frac{u}{2}\nabla_AY^A}$ always appears on the right hand side, and we may subtract it in the definition.  We add a bar in $\bar{\mathcal{M}}_Y$ to distinguish with the flux operator $\mathcal{M}_Y$. There is no guarantee that the $\dot{N}_A$ is  proportional to the angular momentum flux operator $M_A(u,\Omega)$. 

Formally, the supertranslation and superrotation generators constructed from the flux operators are rather similar to the extended BMS charge operators \eqref{gstcharge} and \eqref{gsrcharge}. However, the Bondi mass and angular momentum aspects are ambiguous since we may add total time derivative terms in their definitions. In the following, we will show that by adding suitable counterterms at $\mathcal{I}^+$, one may relate the smeared operators defined in previous sections to the extended BMS charge operators in \cite{2020JHEP...10..116C}. 


\subsection{Scalar theory}
For a real massless scalar field coupled to gravity with  fall-off condition \eqref{falloffscalar},
 we find the leading order terms of stress tensor
\bea 
t_{uu}=\dot\Sigma^2,\qquad  t_{uA}=\dot\Sigma \nabla_A\Sigma,\qquad  t_{AB}=-\gamma_{AB}\dot\Sigma \Sigma
\eea from the expression \eqref{scalarstress}. To find the BMS charge in the flat space limit, we turn off the gravitational field.  Then the constraint equations \eqref{bondimass} and \eqref{bondiangular} become
\bea 
\dot M&=&-4\pi t_{uu}\ =\ -4\pi\dot\Sigma^2,\\
\dot{{N}}_A&=&-8\pi (t_{uA}+\frac{1}{8}\gamma^{BC}\nabla_A t_{BC})+4\pi u \nabla_A t_{uu}\nn\\&=&-8\pi (\dot\Sigma\nabla_A\Sigma-\frac{1}{4}\nabla_A(\dot\Sigma\Sigma)-\frac{u}{2}\nabla_A\dot\Sigma^2).\label{bondiangularas}
\eea 
Substituting these into the generalized BMS charges, we find 
\bea 
\bar{\mathcal{T}}_f&=&\int du d\Omega f(u,\Omega)\dot\Sigma^2,\\ 
\bar{\mathcal{M}}_Y&=&\int du d\Omega Y^A(u,\Omega)[\dot\Sigma\nabla_A\Sigma-\frac{1}{4}\nabla_A(\dot\Sigma\Sigma)].
\eea 
Interestingly, the flux $\bar{\mathcal{T}}_f$ is exactly the supertranslation flux $\mathcal{T}_f$ defined in \cite{Liu:2022mne}. The flux $\bar{\mathcal{M}}_Y$ is related to the  flux operator $\mathcal{M}_Y$ by 
\bea 
\bar{\mathcal{M}}_Y=\mathcal{M}_Y+\frac{1}{8}\mathcal{Q}_{\nabla_A\dot Y^A}.
\eea The operator $\mathcal{Q}_h$ has also been defined in \cite{Liu:2022mne} as
\bea 
\mathcal{Q}_h=\int du d\Omega h(u,\Omega)\Sigma^2.
\eea When $\dot Y=0$, the operator $\mathcal{Q}$ is absent and then we find 
\be 
\mathcal{M}_Y=\bar{\mathcal{M}}_Y.
\ee 

We note that the term $\nabla_A(\dot\Sigma\Sigma)$ in \eqref{bondiangularas} is actually a  total time derivative, 
\be \nabla_A(\dot\Sigma\Sigma)=\frac{1}{2}\frac{d}{du}\nabla_A\Sigma^2.
\ee One may modify the definition of $N_A$ by a further shift 
\bea 
N_A^{\text{re}}=N_A+N_A^{\text{c.t.}},
\eea where 
\be 
N_A^{\text{c.t.}}=8\pi \alpha \Sigma\nabla_A\Sigma.
\ee The coefficient $\alpha$ is not fixed so far.
We may justify this counterterm in another way. We assume the counterterm $N_A^{\text{c.t.}}$ to be local, and thus it can only depend on the field $\Sigma$ and its various derivatives. From dimensional analysis, the dimensions of various local operators are 
\bea 
[M]=1,\quad [N_A]=0,\quad [\Sigma]=0,\quad [\dot\Sigma]=1,\quad [\nabla_A\Sigma]=0.
\eea A candidate counterterm should be 
\bea 
N_A^{\text{c.t.}}=8\pi \alpha \Sigma \nabla_A\Sigma.
\eea There may be other terms such as $\Sigma^2\nabla_A\Sigma,\Sigma\nabla_A\nabla_B\nabla^B\Sigma$ from dimensional analysis. However, these terms are messy. We will impose two conditions to the counterterms 
\begin{enumerate}
\item The number of the field $\Sigma$ in the counterterm should be no more than 2
\bea 
\#(\Sigma)\le 2.\label{numscalar}
\eea The number of the field in $\Sigma^2\nabla_A\Sigma$ is 3, therefore we do not use it.
\item  The number of the derivatives $\nabla_A$ should be no more than 2
\be 
\#(\nabla_A)\le 2.\label{numderi}
\ee This condition rules out the terms like $\Sigma\nabla_A\nabla_B\nabla^B\Sigma$.
\end{enumerate} Therefore, we find a one-parameter family of the Bondi angular momentum and 
\bea 
\dot N_A^{\text{re}}=-8\pi [\dot\Sigma\nabla_A\Sigma-\frac{1+4\alpha}{4}\nabla_A(\dot\Sigma\Sigma)-\frac{u}{2}\nabla_A\dot\Sigma^2].
\eea 
The corresponding superrotation charge is 
\bea 
\bar{\mathcal{M}}^{\text{re}}_Y(\alpha)=-\frac{1}{8\pi}\int du d\Omega Y^A(u,\Omega)\dot{N}^{\text{re}}_A-\bar{\mathcal{T}}_{f=\frac{1}{2}u\nabla_AY^A}=\mathcal{M}_Y(\lambda)
\eea where $\mathcal{M}_Y(\lambda)$ has been defined in \cite{Liu:2022mne} as
\bea  
\mathcal{M}_Y(\lambda)=\int du d\Omega Y^A(u,\Omega)( \lambda \dot\Sigma \nabla_A\Sigma-(1-\lambda)\Sigma\nabla_A\dot\Sigma).
\eea The relation between $\lambda$ and $\alpha$ is
\bea 
\lambda=\frac{3}{4}-\alpha.
\eea In \cite{Liu:2022mne}, $\lambda=\frac{1}{2}$ is singled out by the orthogonality condition 
\be 
\langle \mathcal{T}_f\mathcal{M}_Y\rangle=0.\label{orth}
\ee It is equivalent to adding a counterterm with $\alpha=\frac{1}{4}$. The lesson from the scalar theory is that one may add  total time derivative terms to both \eqref{bondimass} and \eqref{bondiangular}. This is equivalent to modifying the Bondi mass aspect and Bondi angular momentum aspect by counterterms.

\subsection{Electromagnetic theory}
For a vector field $a_\mu$ coupled to gravity, one imposes the following fall-off condition\footnote{We choose the radial gauge $a_r=0$ for convenience.}
\bea 
a_u&=&\frac{A_u(u,\Omega)}{r}+\mathcal{O}\left(\frac{1}{r^2}\right),\\ 
a_A&=&A_A(u,\Omega)+\mathcal{O}\left(\frac{1}{r}\right).
\eea The stress tensor of free electromagnetic theory is 
\bea 
T_{\mu\nu}=f_{\mu\rho}f^{\ \rho}_\nu-\frac{1}{4}g_{\mu\nu} f_{\rho\sigma}f^{\rho\sigma},
\eea where the antisymmetric tensor $f_{\mu\nu}$ reads
\bea 
f_{\mu\nu}=\partial_\mu a_\nu-\partial_\nu a_\mu.
\eea 
We find the following leading terms of stress tensor
\bea 
t_{uu}=\gamma^{AB}\dot A_A\dot A_B,\qquad t_{uA}=A_u \dot A_A+\dot{A}^C(\nabla_AA_C-\nabla_CA_A),\qquad t_{AB}=0
\eea and thus
\bea 
\dot M&=&-4\pi \dot{A}_A\dot{A}^A,\\ 
\dot N_A&=&-8\pi t_{uA}-u\nabla_A\dot M.
\eea Now it is easy to check 
\bea 
\bar{\mathcal{T}}_f=\mathcal{T}_f.
\eea The superrotation flux is 
\bea 
\bar{\mathcal{M}}_Y=\int du d\Omega Y^A t_{uA}=\int du d\Omega Y^A[A_u \dot A_A+\dot{A}^C(\nabla_AA_C-\nabla_CA_A)].
\eea 
From the equation of motion, one can determine $A_u$
\bea 
\dot A_u=\nabla_A\dot A^A\quad\Rightarrow \quad A_u=\nabla_A A^A+\varphi(\Omega).
\eea
Therefore, we find 
\bea 
\bar{\mathcal{M}}_Y-\mathcal{M}_Y=\frac{1}{2}\int du d\Omega Y^A \frac{d}{du}[A^B\nabla^CA^D P_{ABCD}]+\int du d\Omega Y^A \dot{A}_A \varphi(\Omega).
\eea We may add local counterterms 
\bea 
N_A^{\text{c.t.}}=8\pi[\frac{1}{2}A^B\nabla^CA^DP_{ABCD}+A_A\varphi(\Omega)]
\eea to the Bondi angular momentum aspect and get
\bea 
N_A^{\text{re}}=N_A+N_A^{\text{c.t.}}.
\eea The corresponding superrotation flux becomes 
\bea 
\bar{\mathcal{M}}^{\text{re}}_Y=-\frac{1}{8\pi}\int du d\Omega Y^A N_A^{\text{re}}-\bar{\mathcal{T}}_{f=\frac{1}{2}u\nabla_A Y^A}=\mathcal{M}_Y.
\eea 
More generally, we may add the counterterm below
\be 
N_A^{\text{c.t.}}(\alpha)=8\pi\varphi(\Omega)A_A+8\pi\alpha A^B\nabla^CA^D P_{ABCD}.
\ee As a consequence, 
the superrotation flux becomes 
\bea 
\bar{\mathcal{M}}^{\text{re}}_Y(\alpha)&=&\mathcal{M}_Y(\lambda),\qquad \lambda=1-\alpha.
\eea 

\subsection{Gravitational theory}
For pure Einstein gravity, the constraint equations are
\bea 
\dot{M}&=&-\frac{1}{8G}\dot{C}_{AB}\dot{C}^{AB}+\frac{1}{4G}\nabla_A\nabla_B \dot{C}^{AB},\label{bondimass2}\\
\dot{\bar{N}}_A&=&\nabla_AM+\frac{1}{4G}\nabla^B(\nabla_A\nabla^CC_{BC}-\nabla_B\nabla^C C_{AC})\nn\\&&+\frac{1}{4G}\nabla_B(\dot{C}^{BC}C_{AC})+\frac{1}{2G}\nabla_B\dot{C}^{BC} C_{AC}.\label{bondiangular2}
\eea 
On the right hand side of the first equation, the second term which is linear in the shear tensor relates to the memory effect. 
We find the supermomentum charge 
\bea 
\bar{\mathcal{T}}_f=\mathcal{T}_f+\frac{1}{16\pi G}\int du d\Omega f(u,\Omega)\nabla_A\nabla_B\dot{C}^{AB}.
\eea 
We may add a local counterterm to modify the Bondi mass aspect
\be 
M^{\text{re}}=M+M^{\text{c.t.}},\qquad M^{\text{c.t.}}=-\frac{1}{4G}\nabla_A\nabla_BC^{AB}.\label{ctM}
\ee With this modification, the corresponding supermomentum charge is exactly the supertranslation generator defined in previous section
\be 
\bar{\mathcal{T}}_f=-\frac{1}{4\pi}\int du d\Omega f(u,\Omega) M^{\text{re}}=\frac{1}{32\pi G}\int du d\Omega f(u,\Omega)\dot{C}_{AB}\dot{C}^{AB}=\mathcal{T}_f.
\ee 

Now we turn to the Bondi angular momentum aspect. With the definition \eqref{shiftN}, we find \cite{2020JHEP...10..116C}
\bea 
\dot{N}_A&=&\frac{u}{8G}\nabla_A(\dot{C}_{BC}\dot{C}^{BC})+\frac{1}{4G}H_A(C,\dot C)\nn\\
&&-\frac{u}{4G}[\nabla_A\nabla_B\nabla_C\dot{C}^{BC}+\nabla_B\nabla_A\nabla_C\dot{C}^{BC}-\nabla^B\nabla_B\nabla^C\dot{C}_{AC}].
\eea Then the extended superrotation flux is 
\bea 
\bar{\mathcal{M}}_Y=\mathcal{M}_Y+\text{linear terms in the news}.
\eea $\mathcal{M}_Y$ is quadratic in the news or shear, while the terms linear in the news are soft part which corresponds to the memory effect. The linear terms may be canceled by adding a counterterm
\bea 
N^{\text{c.t.}}_A=\frac{1}{4G}[\nabla_A\nabla_B\nabla_C \mathcal{C}^{BC}+\nabla_B\nabla_A\nabla_C\mathcal{C}^{BC}-\nabla^B\nabla_B\nabla^C\mathcal{C}_{AC}]
\eea with 
\bea 
\mathcal{C}_{AB}(u,\Omega)=\int^u du' u' \dot{C}_{AB}(u',\Omega)
\eea 
Unlike the scalar and vector theory, the counterterm is non-local. However, we note that the operator $\mathcal{C}_{AB}$ is associated to the spin memory effect. This is similar to the one in \eqref{ctM} where the counterterm is associated with the displacement memory effect.

\subsection{Further comparisons}
As a matter of fact, in the above comparison with \cite{2020JHEP...10..116C} we turn off all the gravitational fields for the scalar and vector theories, and it turns out that only the superrotation flux needs to be renormalized since we single out a particular flux by virtue of the orthogonality condition \eqref{orth} at quantum level from the family of classically equivalent fluxes. For gravitational theory, we turn on gravitational fields and turn off matter fields. 
To agree with our previous results, we remove the soft parts through counterterms in both supermomentum and superrotation fluxes. 

However, as shown in \cite{Compere:2018ylh,Donnay:2021wrk,Donnay:2022hkf}, one can make a separation of hard and soft variables in phase space. Boundary terms can be added into the Einstein-Hilbert action, and the renormalized boundary symplectic structure could be divided into hard and soft parts. They give hard and soft surface charges on a spatial section of $\mathcal{I}^+$, and also  hard and soft fluxes on $\mathcal{I}^+$. 

Such a separation of hard and soft parts has been justified through leading soft graviton theorem (for  supermomentum flux) and subleading soft graviton theorem (for superrotation flux) \cite{Compere:2018ylh}.  Moreover, it has been shown that the total charges form a charge algebra under modified Lie bracket (see (5.68) in \cite{Compere:2018ylh}), while in \cite{Donnay:2022hkf}, the authors have shown that the fluxes of hard and soft parts can generate the transformations on the 
phase space, and form a representation of extended BMS algebra (see (3.24) in \cite{Donnay:2022hkf} or (5.10) in \cite{Donnay:2021wrk}), respectively. 

We have further compared our fluxes with the ones in \cite{Donnay:2022hkf}. When $f$ and $Y$ are time independent, our fluxes $\mathcal{T}_f$ and $\mathcal{M}_Y$ agree with the hard parts of (3.15) and (3.16) in \cite{Donnay:2022hkf} whose authors renormalize the phase space by separating the hard/soft variables and adding boundary terms to the action, and gain integrable fluxes through the formula $\delta H_{\bm\xi}=i_{\bm\xi}\bm\Omega$. We have also used this equation to check our fluxes in section \ref{canochargegrav}. To make the hard part integrable, we use the notion of covariant variation $\slashed\delta_YC_{AB}$ which has been proposed in \cite{Liu:2023qtr} to make the superrotation variation for electromagnetic field compatible with boundary metric $\gamma_{AB}$ and make the corresponding superrotation fluxes integrable. We should emphasize that although our processing (modifying the variation) is different from renormalizing the  symplectic form, these two methods give the same hard fluxes.

There is an important property about the integrability of the BMS charge. As stated in \cite{Wald:1999wa,Anninos:2010zf,Compere:2020lrt}, when using $\int_{S^2} {\bf k}_{\bm\xi}$ to construct the surface charge $Q_{\bm\xi}$ on a spatial section of $\mathcal{I}^+$, the non-integrable part is the flux of the integrable charge. In the last comment of section 2 in \cite{Compere:2020lrt}, the authors take a particular symmetry generator $\p_t$ in the charge algebra such that the time derivative of integrable charge $\frac{d}{dt}Q_{\bm\xi}[\phi]$ is exactly represented by the non-integrable part as $-\Xi_{\p_t}[\delta_{\bm\xi}\phi;\phi]$. If integrating $\frac{d}{dt}Q_{\bm\xi}[\phi]$ with respect to time, then one will get the (integrated) flux on $\mathcal{I}^+$ which is also the integration of $-\Xi_{\p_t}[\delta_{\bm\xi}\phi;\phi]$. These integrated fluxes also agree with our results. 

In summary, there are mature treatments in the literature for these issues and we refer the interested readers to \cite{Freidel:2021fxf,Compere:2018ylh,Compere:2019bua,Barnich:2021dta,Donnay:2021wrk,Donnay:2022hkf,2020JHEP...10..116C,Compere:2020lrt,Ruzziconi:2020cjt} and the references therein.

\section{Conclusion and discussion}\label{cd}
In this paper, we have reduced the linearized 
gravity theory in Minkowski spacetime to future null infinity $\mathcal{I}^+$.  The boundary tensor theory is characterized by the shear tensor $C_{AB}$ with a non-trivial symplectic form.  The ten Poincar\'e fluxes are totally determined by the shear tensor. We have defined the flux operators and interpreted them as supertranslation and superrotation generators. As in the electromagnetic theory, one should define a covariant variation to identify the superrotation generators. The flux operators do not form a closed algebra in general. There is a truncated Lie algebra \eqref{Tf1Tf2}-\eqref{OgOh}, if a gravitational duality operator $\mathcal{O}_g$ is included, and the parameters satisfy $\dot Y=\dot g=0$. The infinite dimensional algebra is isomorphic to the one in the electromagnetic theory. We provide three different ways to understand the flux operators $\mathcal{T}_f, \mathcal{M}_Y$ and the duality operator $\mathcal{O}_g$, which will be compared in the following.  
\begin{itemize}
    \item Physical approach. This is also the main method used in our previous paper\cite{Liu:2022mne,Liu:2023qtr}. In this way, we find the Poincar\'e fluxes as well as the helicity flux corresponding to gravitational duality transformation from the  conserved currents in the bulk, and thus read out the flux density operators. To preserve the time and angular dependence information in the flux density operators, one may try to transform the flux density operators to its (generalized) Fourier space and define the corresponding smeared operators $\mathcal{T}_f, \mathcal{M}_Y$ and $\mathcal{O}_g$. The test functions $f,Y, g$ are assumed to be time and angular dependent. After calculating the lengthy commutators among these operators, one finds that it is necessary to require the following conditions
    \be 
    f=f(u,\Omega),\qquad Y^A=Y^A(\Omega),\qquad g=g(\Omega),\label{restriction}
    \ee 
    if we want a closed Lie algebra. In this approach, the physical meaning of the operators are clear. 
    
    \item Hamiltonians from boundary theory.  The flux operators can also be realized as Hamiltonians from the boundary Carrollian field theory. In this approach, the boundary theory is determined by a solution space which should satisfy the boundary constraints. The solution space is equipped with a symplectic form which could be used to obtain the Hamiltonian through the formula \eqref{vacharge}. The operators $\mathcal{T}_f$ and $\mathcal{M}_Y$ are identified with the Hamiltonians corresponding to Carrollian diffeomorphisms. At the same time, the operator $\mathcal{O}_g$ is identified with the Hamiltonian of the extended gravitational duality transformation at $\mathcal{I}^+$. In this approach, the condition \eqref{restriction} is found automatically by requiring the Hamiltonian to be integrable. We could also obtain a general formula \eqref{Hzeta} which may be valid for general Carrollian field theories. 
    
    \item BMS charges from bulk theory. Though the flux operators $\mathcal{T}_f, \mathcal{M}_Y$ and $\mathcal{O}_g$ are obtained in linearized gravity, we could find their relations to the BMS charges in fully nonlinear Einstein gravity. The identification is not straightforward, and one need add counterterms to the Bondi mass and angular momentum aspects. The counterterms are local for the scalar and vector theory, while they could  be non-local for the gravitational theory. It is not clear whether one can find a unique way to add the counterterms at this moment.
\end{itemize}
There are various open questions that deserve further study.
\begin{itemize}
\item Boundary theory in asymptotically flat spacetime. The starting point of our work is to embed the boundary theory at $\mathcal{I}^+$ to four dimensional spacetime in which the field theory is well known. However, there should be an intrinsic way to define the boundary theory from the Carrollian diffeomorphism of $\mathcal{I}^+$.
\item Hamiltonians.  We could define the Hamiltonians from the symplectic form of the boundary theory. The Hamiltonians are integrable for GSTs and SSRs as well as SSDTs which could form a closed Lie algebra. On the other hand, they fail to be integrable for GSRs and GSDTs. This is consistent with the fact that the flux operators corresponding to GSRs and GSDTs would lead to non-local terms in the commutators. There may be deep connections between the non-integrability and non-local terms. 
    \item Subleading terms and interactions. Our work mainly focuses on the leading terms in the fall-off conditions and they are related to radiative modes in the bulk. The radiative modes are free from EOM in the boundary theory which is universal for general bulk theories. Namely, the boundary theories could be the same at the leading order for different bulk theories. Therefore, to distinguish different bulk theories, one may delve into the subleading terms in the fall-off conditions. These terms are related to the radiative modes and the coupling constants  through the constraint equations.
\item Fall-off conditions. We derive the flux densities corresponding to the Killing symmetry, and then use these densities to construct flux operators related to the Carrollian diffeomorphism. However, it remains a problem how to extend the Carrollian diffeomorphism to the bulk and how this extension will affect the fall-offs and the solution space.
\end{itemize}


\vspace{10pt}
{\noindent \bf Acknowledgments.} 
The work of J.L. is supported by NSFC Grant No. 12005069.
\appendix

\section{Properties of the vectors \texorpdfstring{$n^\mu,\bar{n}^\mu, Y_A^\mu$}{}}\label{ckvmink}
The null vectors $n^\mu$ and $\bar n^\mu $ are defined as 
\be 
n^\mu=(1,n^i),\qquad \bar{n}^\mu=(-1,n^i).
\ee The vectors $Y^\mu_A$ are defined as 
\bea Y^A_\mu=-\nabla^A n_\mu=-\nabla^A\bar n_\mu.
\eea  The Greek indices $\mu,\nu,\cdots$ are raised by $\eta^{\mu\nu}$ while the Latin indices $A,B,\cdots$ are raised by $\gamma^{AB}$. We may use $n^\mu$ and $Y^\nu_A$ to define the conformal Killing vectors of the unit sphere 
\bea 
Y^{\mu\nu}_A=Y^\mu_A n^\nu-Y^\nu_A n^\mu.
\eea $Y_{\mu\nu}^A$ is antisymmetric  
\be 
Y_{\mu\nu}^A=-Y^A_{\nu\mu}.
\ee Its $0i$ components are exactly the strictly conformal Killing vectors
\be 
Y_{0i}^A=Y_i^A 
\ee and $ij$ components are the conformal Killing vectors $Y_{ij}^A$ which are defined in \cite{Liu:2022mne}. We list the  properties in the following.
\begin{enumerate}
    \item Orthogonality
    \be 
   n^\mu n_\mu=\bar{n}^\mu\bar{n}_\mu=0,\quad  n^\mu Y_\mu^A=\bar{n}^\mu Y_\mu^A=0,\quad n^\mu\bar{n}_\mu=2,\quad Y_\mu^A Y_\nu^B\eta^{\mu\nu}=\gamma^{AB}
    \ee 
    \item Completeness
    \be \frac{1}{2}(n_\mu\bar{n}_\nu+n_\nu\bar{n}_\mu)+Y_\mu^AY^B_\nu\gamma_{AB}=\eta_{\mu\nu}.\label{completeY}
    \ee 
    \item The identities involve covariant derivatives of $Y_\mu^A$ 
    \bea
   && \nabla_AY_\mu^A=n_\mu+\bar{n}_\mu,\quad \quad n^\alpha\nabla_AY^B_\alpha=\delta^B_A,\\
   &&Y_\alpha^B\nabla_AY^\alpha_C=0,\qquad \nabla_AY_B^\mu-\nabla_BY_A^\mu=0.
    \eea
    \item The covariant derivative of $Y_{\mu\nu}^A$ takes the form
\bea n^\mu\bar{n}^\nu-n^\nu\bar{n}^\mu=-\nabla_AY^{\mu\nu A}.
\eea
\item The identity involves two CKVs
\bea 
Y_{\rho\sigma}^AY_{\mu\nu}^BQ_{AB}^{\ \ \ CD}+\frac{1}{8}\epsilon_{\mu\nu}^{\hspace{11pt}\alpha\beta}(Y_{\alpha\beta}^CY_{\rho\sigma}^D+Y_{\alpha\beta}^DY_{\rho\sigma}^C)+\frac{1}{8}\epsilon_{\rho\sigma}^{\hspace{11pt}\alpha\beta}(Y_{\alpha\beta}^CY_{\mu\nu}^D+Y_{\alpha\beta}^DY_{\mu\nu}^C)=0.\label{2ymunu}
\eea 
\item The normal vector $n^i$ may be lifted to a four-vector $(0,n^i)$ which is the average of the null vectors $n^\mu$ and $\bar n^\mu$
\be 
(0,n^i)=\frac{1}{2}(n^\mu+\bar{n}^\mu).\label{ave}
\ee  Similarly, we can express the four-vector $(1,0)$ as the difference between the null vectors
\begin{align}
    (1,0)=\frac{1}{2}(n^\mu-\bar{n}^\mu).
\end{align}

\end{enumerate}

There are more identities involving more than one normal vector $n^\mu$ and $Y^A_\mu$. To simplify notation, we define the following three tensors 
\bea 
&&N_{\mu\nu}=n_\mu n_\nu,\\
&&U^A_{\mu\nu}=\frac{1}{2}(n_\mu Y_\nu^A+n_\nu Y_\mu^A),\\
&&V^{AB}_{\mu\nu}=\frac{1}{2}(Y_\mu^A Y^B_\nu+Y_\nu^A Y^B_\mu).
\eea They are symmetric under the interchange of the indices $\mu$ and $\nu$
\bea 
N_{\mu\nu}=N_{\nu\mu},\qquad U^A_{\mu\nu}=U^A_{\nu\mu},\qquad V^{AB}_{\mu\nu}=V^{AB}_{\nu\mu},
\eea 
and are transverse to the vector $n^\mu$
\bea 
N_{\mu\nu}n^\nu=0,\qquad U^A_{\mu\nu}n^\nu=0,\qquad V^{AB}_{\mu\nu}n^\nu=0.
\eea 
Moreover, one can find the following trace
\bea 
&&N^\mu_{\ \mu}=0,\qquad U^{A}_{\mu\nu}\eta^{\mu\nu}=0,\qquad V^{AB}_{\mu\nu}\eta^{\mu\nu}=\gamma^{AB},\\ &&V^{AB}_{\mu\nu}\gamma_{AB}=\eta_{\mu\nu}-\frac{1}{2}(n_\mu\bar{n}_\nu+n_\nu\bar{n}_\mu).
\eea 
We can also compute their squares 
\bea 
&&N_{\mu\nu}N^{\mu\nu}=0,\qquad U_{\mu\nu}^A U^{B\mu\nu}=0,\\ &&V^{AB}_{\mu\nu}V^{CD\mu\nu}=\frac{1}{2}(\gamma^{AC}\gamma^{BD}+\gamma^{AD}\gamma^{BC}).
\eea
These tensors are orthogonal to each other
\bea 
N^{\mu\nu}U_{\mu\nu}^A=N^{\mu\nu}V^{AB}_{\mu\nu}=U^A_{\mu\nu}V^{\mu\nu BC}=0.
\eea  
Their products with the vector $Y_\mu^A$ read
\bea 
&&N_{\mu\nu}Y^{\nu A}=0,\qquad U^A_{\mu\nu}Y^{\nu B}=\frac{1}{2}\gamma^{AB}n_\nu,\\
&&V^{AB}_{\mu\nu}Y^{\nu C}=\frac{1}{2}(\gamma^{BC}Y_\mu^A+\gamma^{AC}Y_\mu^B).
\eea 
We can also find the following products
\bea 
&&N_{\mu\alpha}N^{\alpha}_{\ \nu}=N_{\mu\alpha} U^{\alpha A}_{\hspace{10pt} \nu}= N_{\mu\alpha}V^{\alpha AB}_{\hspace{15pt} \ \nu}=0,\\
&& U^A_{\mu\alpha}U^{\alpha B}_{\hspace{10pt}\nu}=\frac{1}{4}\gamma^{AB}n_\mu n_\nu,\\ && U^A_{\mu\alpha}V^{\alpha BC}_{\hspace{20pt}\nu}=\frac{1}{4}(n_\mu Y^C_\nu\gamma^{AB}+n_\mu Y_\nu^B\gamma^{AC}),\\&& V^{AB}_{\mu\alpha}V^{\alpha CD}_{\hspace{20pt}\nu}=\frac{1}{4}(\gamma^{BC}Y_\mu^AY_\nu^D+\gamma^{BD}Y_\mu^AY_\nu^C+\gamma^{AC}Y_\mu^BY_\nu^D+\gamma^{AD}Y_\mu^BY^C_\nu).
\eea For the derivatives of the symmetric tensors $N_{\mu\nu}, U^A_{\mu\nu}$ and $V^{AB}_{\mu\nu}$, we find 
\bea 
&&N^{\mu\nu}\nabla_A N_{\mu\nu}=N^{\mu\nu}\nabla_A U^B_{\mu\nu}=N^{\mu\nu}\nabla_A V^{BC}_{\mu\nu}=U^{A\mu\nu}\nabla_B N_{\mu\nu}=0,\\
&&U^{\mu\nu A}\nabla_B U^C_{\mu\nu}=V^{\mu\nu AB}\nabla_C N_{\mu\nu}=V^{\mu\nu AB}\nabla_C V^{DE}_{\mu\nu}=0,\\ 
&& U^{\mu\nu A}\nabla_B V_{\mu\nu CD}=-V^{\mu\nu}_{CD}\nabla_BU^A_{\mu\nu}=\frac{1}{2}(\gamma^A_C\gamma_{BD}+\gamma^A_D\gamma_{BC}).\eea 
When they are contracted with $n_\mu$ or $Y_\mu^A$, one can find 
\bea 
&& n^\beta\nabla_A N_{\beta\mu}=0,\\ 
&& n^\beta\nabla_A U^B_{\beta\mu}=\frac{1}{2}\gamma^B_A n_\mu,\\ 
&& n^\beta\nabla_A V^{BC}_{\beta\mu}=\frac{1}{2}(\gamma^B_A Y_\mu^C+\gamma^C_A Y_\mu^B),\\ 
&& Y_D^\beta \nabla_A N_{\beta\mu}=-\gamma_{AD}n_\mu,\\
&& Y_D^\beta\nabla_A U^B_{\beta\mu}=-\frac{1}{2}(\gamma_{AD}Y_\mu^B+\gamma^B_D Y_{\mu A}),\\ 
&& Y^\beta_D\nabla_A V^{BC}_{\beta\mu}=\frac{1}{2}(\gamma^B_D \nabla_A Y_\mu^C+\gamma^C_D\nabla_AY_\mu^B).
\eea 
More identities  are listed as follows 
\bea 
&& N^{\mu\alpha}\nabla_A N_{\alpha\nu}=0,\\ 
&& N^{\mu\alpha}\nabla_A U^B_{\alpha\nu}=\frac{1}{2}\gamma^B_A n^\mu n_\nu,\\ 
&& N^{\mu\alpha}\nabla_A V^{BC}_{\alpha\nu}=\frac{1}{2}\gamma^B_{A}n^\mu Y_\nu^C+\frac{1}{2}\gamma^C_A n^\mu Y_\nu^B,\\
&& U^{\mu\alpha A}\nabla_B N_{\alpha\nu}=-\frac{1}{2}\gamma^A_B n^\mu n_\nu,\\
&& U^{\mu\alpha A}\nabla_B U^C_{\alpha\nu}=\frac{1}{4}(-\gamma^A_B n^\mu Y_\nu^C-\gamma^{AC}n^\mu Y_{\nu B}+\gamma_B^C n_\nu Y^{\mu A}),\\
&& U^{\mu\alpha A}\nabla_B V^{CD}_{\alpha\nu}=\frac{1}{4}(\gamma_B^C Y^{\mu A}Y_\nu^D+\gamma_B^D Y^{\mu A}Y_\nu^C+\gamma^{AC}n^\mu \nabla_B Y_\nu^D+\gamma^{AD}n^\mu \nabla_B Y_\nu^C),\\
&& V^{\mu\alpha AB}\nabla_C N_{\alpha\nu}=-\frac{1}{2}(\gamma^B_C Y^{\mu A}n_\nu+\gamma^A_C Y^{\mu B}n_\nu),\\
&& V^{\mu\alpha AB}\nabla_C U^D_{\alpha\nu}=-\frac{1}{4}(\gamma^B_C Y^{\mu A}Y_\nu^D+\gamma^{BD}Y^{\mu A}Y_{\nu C}+\gamma^A_C Y^{\mu B}Y_\nu^D+\gamma^{AD}Y^{\mu B}Y_{\nu C}),\\
    && V^{\mu\alpha AB}\nabla_C V^{DE}_{\alpha\nu}=\frac{1}{4}(\gamma^{BD}Y^{\mu A}\nabla_C Y_\nu^E+\gamma^{AD}Y^{\mu B}\nabla_C Y_\nu^E+\gamma^{BE}Y^{\mu A}\nabla_CY_\nu^D+\gamma^{AE}Y^{\mu B}\nabla_CY_\nu^D).\nn\\
\eea 
The above equations lead to the following 
\bea 
&& Y_\mu^A N^{\mu\alpha}\nabla_A N_{\alpha\nu}=0,\\ 
&& Y_\mu^A N^{\mu\alpha}\nabla_A U^B_{\alpha\nu}=0,\\ 
&& Y_\mu^A N^{\mu\alpha}\nabla_A V^{BC}_{\alpha\nu}=0,\\
&& Y_\mu^B U^{\mu\alpha A}\nabla_B N_{\alpha\nu}=0,\\
&& Y_\mu^BU^{\mu\alpha A}\nabla_B U^C_{\alpha\nu}=\frac{1}{4}\gamma^{AC} n_\nu,\\
&& Y_\mu^BU^{\mu\alpha A}\nabla_B V^{CD}_{\alpha\nu}=\frac{1}{4}(\gamma^{AC}Y_\nu^D+\gamma^{AD}Y_\nu^C),\\
&& Y^C_\mu V^{\mu\alpha AB}\nabla_C N_{\alpha\nu}=-\gamma^{AB}n_\nu,\\
&& Y^C_\mu V^{\mu\alpha AB}\nabla_C U^D_{\alpha\nu}=-\frac{1}{4}(2\gamma^{AB}Y_\nu^D+\gamma^{BD}Y_\nu^A+\gamma^{AD}Y_\nu^B),\\
    && Y^C_\mu V^{\mu\alpha AB}\nabla_C V^{DE}_{\alpha\nu}=\frac{1}{4}(\gamma^{BD}\nabla^A Y_\nu^E+\gamma^{AD}\nabla^B Y_\nu^E+\gamma^{BE}\nabla^AY_\nu^D+\gamma^{AE}\nabla^BY_\nu^D).
\eea 

\section{Higher rank tensors in Minkowski spacetime}
\subsection{Properties of the tensor \texorpdfstring{$L^{\mu_1\mu_2\cdots\mu_6}$}{}}\label{Lpro}
The tensor $L^{\mu_1\mu_2\cdots\mu_6}$ defined in Minkowski spacetime has the following properties. 
\begin{enumerate}
    \item Symmetries. The tensor $L^{\mu_1\mu_2\cdots\mu_6}$ is invariant under interchange of the second index and the third 
    \be 
    L^{\mu_1\mu_2\mu_3\mu_4\mu_5\mu_6}=L^{\mu_1\mu_3\mu_2\mu_4\mu_5\mu_6}.
    \ee It is also invariant under the interchange of the fifth index and the sixth
    \be 
    L^{\mu_1\mu_2\mu_3\mu_4\mu_5\mu_6}=L^{\mu_1\mu_2\mu_3\mu_4\mu_6\mu_5}.
    \ee
The tensor $L^{\mu_1\mu_2\cdots\mu_6}$ is invariant under the interchange of the first three indices and the last three indices
\bea 
L^{\mu_1\mu_2\mu_3\mu_4\mu_5\mu_6}=L^{\mu_4\mu_5\mu_6\mu_1\mu_2\mu_3}.
\eea 

\item Identities involving normal vector $n^\mu$ and conformal Killing vectors $Y^A_\mu$. 
The following identities can be checked straightforwardly.
\bea 
L^{\mu_1\mu_2\mu_3\mu_4\mu_5\mu_6}N_{\mu_2\mu_3}N_{\mu_4\mu_5}&=&0,\\
L^{\mu_1\mu_2\mu_3\mu_4\mu_5\mu_6}n_{\mu_4}N_{\mu_2\mu_3}V_{\mu_5\mu_6}^{AB}&=&0,\\ 
L^{\mu_1\mu_2\mu_3\mu_4\mu_5\mu_6}n_{\mu_4}N_{\mu_5\mu_6}U^A_{\mu_2\mu_3}&=&0,\\
L^{\mu_1\mu_2\mu_3\mu_4\mu_5\mu_6}n_{\mu_4}U^A_{\mu_2\mu_3}U^B_{\mu_5\mu_6}&=&0,\\ 
L^{\mu_1\mu_2\mu_3\mu_4\mu_5\mu_6}n_{\mu_4}U_{\mu_2\mu_3}^A V^{CD}_{\mu_5\mu_6}&=&0.
\eea We may also need the following identity 
\bea 
\hspace{-15pt}\frac{1}{2}(n_{\mu_1}+\bar{n}_{\mu_1})L^{\mu_1\mu_2\mu_3\mu_4\mu_5\mu_6}n_{\mu_4}V_{\mu_2\mu_3}^{AB}V_{\mu_5\mu_6}^{CD}&=&\frac{1}{2}(\gamma^{AC}\gamma^{BD}+\gamma^{AD}\gamma^{BC})-\gamma^{AB}\gamma^{CD}.
\eea 
\end{enumerate}
 \subsection{Traces of the tensor \texorpdfstring{$S_{\mu_1\mu_2\cdots\mu_6}$}{}}\label{tracesS}
 We use the notation that 
 \bea 
 (ij)(kl)\equiv S_{\mu_1\cdots\mu_6}\eta^{\mu_i\mu_j}\eta^{\mu_k\mu_l},\quad i,j,k,l=1,2,\cdots,6.
 \eea To find the radiation fluxes, we need the various traces of the tensor $S$. The following traces vanish which have been used in the context. 
 \bea 
&& (15)(46)=(14)(56)=(13)(46)=(24)(56)=(23)(45)=(23)(56)=(15)(23)=(14)(23)=0.\nn
 \eea We also need the following nonvanishing traces. 
 \bea 
 && (14)(26)=\frac{d}{du}(H_{\mu_3}^{\ \alpha}H_{\alpha\mu_5}),\\ 
 &&(25)(34)=H_{\mu_1\alpha}\dot{H}^{\alpha}_{\ \mu_6}+\frac{1}{2}n_{\mu_1}\bar{n}^{\mu_3}(\dot{H}_{\mu_2\mu_3}H^{\mu_2}_{\ \mu_6}-H_{\mu_2\mu_3}\dot{H}^{\mu_2}_{\ \mu_6})\nn\\&&\hspace{2cm}+n_{\mu_1}Y^A_{\mu_3}\dot{H}^{\mu_3\mu_5}\nabla_A H_{\mu_5\mu_6}.,\\
 &&(25)(36)=2n_{\mu_1}n_{\mu_4}\dot{H}^{\alpha\beta}\dot{H}_{\alpha\beta}^{(2)}+[n_{\mu_1}n_{\mu_4}+\frac{1}{2}(n_{\mu_1}\bar{n}_{\mu_4}+n_{\mu_4}\bar{n}_{\mu_1})]\dot{H}^{\alpha\beta}H_{\alpha\beta}\nn\\&&\hspace{2cm}+(n_{\mu_1}Y_{\mu_4}^A+n_{\mu_4}Y_{\mu_1}^A)\dot{H}^{\alpha\beta}\nabla_AH_{\alpha\beta},\\
 &&(14)(25)=\frac{d}{du}(H_{\mu_3\alpha}H_{\mu_6}^{\ \alpha}),\\ 
 &&(15)(24)=\frac{d}{du}(H_{\mu_3\alpha}H^{\alpha}_{\ \mu_6}).
 \eea 

\subsection{Various combinations of \texorpdfstring{$H_{\mu\nu}$}{} and \texorpdfstring{$H_{\mu\nu}^{(2)}$}{}}\label{hh2}
To find the radiation fluxes,  we may need the following identities
\bea 
n^\mu H_{\mu\nu}&=&0,\\ 
\bar{n}^\mu H_{\mu\nu}&=&n_\nu(4GM)+Y_\nu^A\nabla^BC_{AB},\\ 
H_{\mu\nu}\eta^{\mu\nu}&=&0,\\ 
H_{\mu\nu}^{(2)}\eta_{\mu\nu}&=&\gamma^{AB}Z_{AB}-2\widetilde{X},\\
H_{\mu\nu}H^{\mu\nu}&=&C_{AB}C^{AB},\\
n^\mu H^{(2)}_{\mu\nu}&=&-\widetilde{X}n_\nu,\\
n^\mu n^\nu H^{(2)}_{\mu\nu}&=&0,\\
\bar{n}^\mu H^{(2)}_{\mu\nu}&=&2(X-\widetilde{X})n_\nu-\widetilde{X}\bar{n}_\nu+Y_\nu^AJ_A,\\
H^{\mu\alpha}H_{\alpha}^{\ \nu}&=&\frac{1}{4}n^\mu n^\nu \nabla^BC_{AB}\nabla^CC^{A}_{\ C}+\frac{1}{2}(n^\mu Y^{\nu C}+n^\nu Y^{\mu C})C^{A}_{\ C}\nabla^BC_{AB}\nn\\&&+Y^{\mu A}Y^{\nu B}C_{A}^{\ C}C_{CB},\\
\dot{H}^{\mu\alpha}H_{\alpha}^{\ \nu}&=&\frac{1}{4}n^\mu n^\nu \nabla^B\dot{C}_{AB}\nabla_CC^{AC}+\frac{1}{2}n^\mu Y^{\nu C}C^A_{\ C}\nabla^B\dot C_{AB}\nn\\&&+\frac{1}{2}n^\nu Y^{\mu C}\dot{C}_{AC}\nabla_BC^{AB}+Y^{\mu C}Y^{\nu D}\dot{C}^{A}_{\ C}C_{AD},\\
\dot{H}^{\alpha\beta}H_{\alpha\beta}&=&\dot{C}_{AB}C^{AB},\\
Y_\mu^A H^{\mu\nu}&=&\frac{1}{2}n^\nu \nabla_C C^{AC}+Y^\nu_C C^{AC},\\
H^{\mu\nu}H^{(2)}_{\mu\nu}&=&Z_{AB}C^{AB}=0,\\
\dot{H}^{\alpha\beta}\nabla_AH_{\alpha\beta}&=&(C_{AC}\nabla_B\dot C^{BC}-\dot{C}_{AC}\nabla_BC^{BC})+\dot{C}_{BC}\nabla_AC^{BC}.\eea


The following two combinations are important for the computation of the angular momentum fluxes.
\bea
n^\nu Y_\alpha^A\dot{H}^{\alpha\beta}\nabla_AH_\beta^{\ \mu}-(\mu\leftrightarrow\nu)&=& \frac{1}{2}Y^{\mu\nu A}(C_{AC}\nabla_B \dot{C}^{BC}-\dot{C}_{AC}\nabla_B C^{BC}+2\dot{C}^{BC}\nabla_BC_{AC})\nn\\&&+\frac{1}{2}(n^\nu \bar{n}^\mu-n^\mu\bar{n}^\nu)\dot{C}^{AB}C_{AB},\\ 
n^\mu\bar{n}^\alpha (\dot{H}_{\alpha\beta}H^{\beta\nu}-H_{\alpha\beta}\dot{H}^{\beta\nu})-(\mu\leftrightarrow\nu)&=&-Y^{\mu\nu A}(C_{AC}\nabla_B\dot{C}^{BC}-\dot{C}_{AC}\nabla_BC^{BC}).
\eea 

\section{Higher rank tensors on \texorpdfstring{$S^2$}{}}\label{rank6}
In this paper, we may use three main higher rank tensors on $S^2$. The rank 4 tensor $P_{ABCD}$ has been defined in the vector theory 
\be 
P_{ABCD}=\gamma_{AB}\gamma_{CD}+\gamma_{AC}\gamma_{BC}-\gamma_{AD}\gamma_{BC}.
\ee The other rank 4 tensor $Q_{ABCD}$ is used to define the duality operator 
\be 
\mathcal{O}_g=\frac{1}{32\pi G}\int du d\Omega  g(u,\Omega)\dot{C}_{AB}C_{CD}Q^{ABCD}
\ee  with 
\bea 
Q_{ABCD}=\frac{1}{4}(\gamma^{BC}\epsilon^{DA}+\gamma^{AC}\epsilon^{DB}+\gamma^{BD}\epsilon^{CA}+\gamma^{AD}\epsilon^{CB}).
\eea 
At last, the rank 6 tensor $P_{ABCDEF}$ is used to define the  angular momentum and center-of-mass flux operators 
\bea 
\mathcal{M}_Y&=&\frac{1}{32\pi G}\int du d\Omega Y^A(u,\Omega)(\dot{C}^{BC}\nabla^DC^{EF}-C^{BC}\nabla^D\dot{C}^{EF})P_{ABCDEF}
\eea with
\bea  
P_{ABCDEF}&=&\frac{1}{4}[\gamma_{AB}(\gamma_{CE}\gamma_{DF}+\gamma_{CF}\gamma_{DE}-\gamma_{CD}\gamma_{EF})+\gamma_{AC}(\gamma_{BE}\gamma_{DF}+\gamma_{BF}\gamma_{DE}-\gamma_{BD}\gamma_{EF})\nn\\&&+\gamma_{AD}(\gamma_{BE}\gamma_{CF}+\gamma_{BF}\gamma_{CE}-\gamma_{BC}\gamma_{EF})-\gamma_{AE}(\gamma_{BD}\gamma_{CF}+\gamma_{BF}\gamma_{CD}-\gamma_{BC}\gamma_{DF})\nn\\&&-\gamma_{AF}(\gamma_{BD}\gamma_{CE}+\gamma_{BE}\gamma_{CD}-\gamma_{BC}\gamma_{DE})-\gamma_{BC}P_{AEFD}+\gamma_{EF}P_{ABCD}]\nn\\&=&\frac{1}{4}(\gamma_{AB}P_{CEFD}+\gamma_{AC}P_{BEFD}+\gamma_{AD}P_{BEFC}-\gamma_{AE}P_{FBCD}-\gamma_{AF}P_{EBCD}\nn\\
&&-\gamma_{BC}P_{AEFD}+\gamma_{EF}P_{ABCD}).
\eea We will study their properties in this appendix.

\subsection{Properties of the rank 4 tensor \texorpdfstring{$P_{ABCD}$}{}} \label{rank4subsec}
The properties of the rank 4 tensor $P_{ABCD}$ are collected in the following. 
Some identities have been obtained in the vector theory. We also add a few new properties which turn out to be useful in this work.
\begin{itemize}
    \item Symmetries
    \bea 
P_{ABCD}=P_{BADC}=P_{BDAC}=P_{DBCA}=P_{CDAB}=P_{DCBA}=P_{ACBD}=P_{CADB}.
\eea \item Traces
\bea 
P^A_{\ ABC}=P^A_{\ BAC}=P_{BC\hspace{5pt}A}^{\hspace{0.4cm} A}=2\gamma_{BC},\quad P^A_{\ BCA}=P_{B\hspace{3pt}AC}^{\ A}=0.
\eea 
\item Fierz identity
\be  
\epsilon^E_{\ B}P_{AECD}+\epsilon^E_{\ D}P_{ABCE}=0.
\ee  This equation comes from the Fierz identity 
\be  
\epsilon_{AB}\gamma_{CD}+\epsilon_{BC}\gamma_{AD}+\epsilon_{CA}\gamma_{BD}=0.
\ee  
\item Product with itself
\be 
\frac{1}{2}P_{ABCD}P_{E\hspace{3pt}F}^{\ B\hspace{3pt}D}=P_{ACEF}.
\ee 
\item The tensor $P_{ABCD}$ can also be written as 
\be 
P_{ABCD}=\gamma_{AC}\gamma_{BD}+\epsilon_{AC}\epsilon_{BD}.
\ee As a consequence, we have
\bea 
P_{ABCD}+P_{ADCB}&=&2\gamma_{AC}\gamma_{BD},\\
P_{ABCD}-P_{ADCB}&=&2(\gamma_{AB}\gamma_{CD}-\gamma_{AD}\gamma_{BC})=2\epsilon_{AC}\epsilon_{BD}.
\eea   \item Square
    \be 
    P_{ABCD}P^{ABCD}=8.
    \ee 
\end{itemize}
\subsection{Properties of the rank 4 tensor \texorpdfstring{$Q_{ABCD}$}{}}
By definition, the rank 4 tensor $Q_{ABCD}$ is constructed from the metric $\gamma_{AB}$ and the Levi-Civita tensor $\epsilon_{AB}$
\bea 
Q^{ABCD}=\frac{1}{4}(\gamma^{BC}\epsilon^{DA}+\gamma^{AC}\epsilon^{DB}+\gamma^{BD}\epsilon^{CA}+\gamma^{AD}\epsilon^{CB}).
\eea 
Its properties are collected below.
\begin{enumerate}
    \item Symmetries
    \bea 
    Q^{ABCD}=Q^{BACD}=Q^{ABDC}=Q^{BADC}.
    \eea 
    \item Traces
    \bea 
    \gamma_{AB}Q^{ABCD}=0,\quad \gamma_{CD}Q^{ABCD}=0,\quad \gamma_{AC}Q^{ABCD}=\epsilon^{DB}.
    \eea 
    \item Antisymmetry
    \be Q^{ABCD}+Q^{CDAB}=0.
    \ee 
    \item Contraction with the rank 6 tensor $P_{ABCDEF}$ 
    \bea 
    Q^{EF}_{\ \ \ BC}P_{AEFDGH}=Q_{GH}^{\ \ \ EF}P_{ABCDEF}.
    \eea 
   As a consequence, one can find
    \bea  
    Q^{BC}_{\ \ \ EF}\rho_{ABCDGH}&=&\frac{1}{2}\gamma_{AD}Q_{GHEF},\\
    Q^{EF}_{\ \ BC}\rho_{AEFDGH}&=&Q_{GH}^{\ \ EF}\rho_{ABCDEF}.
    \eea  
    \item Square
    \be 
    Q_{ABCD}Q^{ABCD}=2.
    \ee 
\end{enumerate}
\subsection{Properties of the rank 6 tensor \texorpdfstring{$P_{ABCDEF}$}{}}
Several properties are listed below. 
\begin{enumerate}
    \item Traces. The tensor $P_{ABCDEF}$ is traceless for the indices $BC$ and $EF$.
\bea 
P_{ABCDEF}\gamma^{BC}=0,\quad P_{ABCDEF}\gamma^{EF}=0.
\eea 
Other useful traces are 
\bea 
P_{ABCDEF}\gamma^{AB}&=&\frac{3}{4}P_{CEFD},\\
P_{ABCDEF}\gamma^{AD}&=&\frac{1}{2}P_{BEFC},\\
P_{ABCDEF}\gamma^{AE}&=&-\frac{1}{4}P_{FBCD},\\
P_{ABCDEF}\gamma^{BD}&=&-\frac{1}{4}P_{CEFA},\\
P_{ABCDEF}\gamma^{ED}&=&\frac{3}{4}P_{FBCA},\\
P_{ABCDEF}\gamma^{BE}&=&\frac{1}{4}(3P_{CAFD}-P_{FACD}).
\eea 
\item Symmetries
\bea 
P_{ABCDEF}=P_{ACBDEF},\quad P_{ABCDEF}=P_{ABCDFE}.
\eea 
\item Algebraic relations
\bea 
P_{ABCDEF}+P_{AEFDBC}=\frac{1}{2}\gamma_{AD}P_{BEFC}\equiv 2\rho_{ABCDEF}.
\eea 
\item Products with $P_{ABCD}$
\bea 
P_{ABCDEF}P^{E\hspace{12pt}F}_{\ \ GH}&=&2P_{ABCDGH},\\
P_{ABCDEF}P^{B\hspace{12pt}C}_{\ \ GH}&=&2P_{AGHDEF}.
\eea
\item Contractions with the shear tensor 
\bea 
\rho_{ABCDEF}C^{EF}&=&\frac{1}{2}\gamma_{AD}C_{BC},\\
P_{AEFDBC}C^{EF}&=&\frac{1}{2}(C_{AB}\gamma_{DC}+C_{AC}\gamma_{DB}+C_{BC}\gamma_{AD}-C_{DC}\gamma_{AB}-C_{DB}\gamma_{AC}).
\eea 
\item Contraction with the tensor $Q_{ABCD}$
\bea 
P_{ABCDEF}Q^{BCEF}=-2\epsilon_{AD}.
\eea 
\item Square
\be 
P_{ABCDEF}P^{ABCDEF}=5.
\ee 
\end{enumerate}

\section{Mode expansion}\label{modeex}
The linearized gravity equation is easily solved by 
imposing de Donder gauge
\begin{align}
    \p_\mu h^{\mu\nu}-\frac{1}{2}\p^\nu h=0,
\end{align}
the PF equation becomes
\begin{align}
  \Box h_{\mu\nu}-\frac{1}{2}\eta_{\mu\nu}\Box h=0 \quad \Rightarrow \quad \Box h_{\mu\nu}=0,
\end{align}
i.e., the relativistic wave equation. We could therefore expand\footnote{We have omitted a normalization factor $\sqrt{32\pi G}$ in this expansion.}
\begin{align}
  h_{{\mu\nu}}(t,\bm x)=&\sum_{\alpha}\int \frac{d^3\bm k}{(2\pi)^3}\frac{1}{\sqrt{2\omega_{\bm k}}}[\epsilon^{*\alpha}_{{\mu\nu}}(\bm k)b_{\alpha,\bm k}e^{-i\omega t+i\bm k\cdot\bm x}+\epsilon^{\alpha}_{{\mu\nu}}(\bm k)b^\dagger_{\alpha,\bm k}e^{i\omega t-i\bm k\cdot\bm x}],\label{hmnexp}
\end{align}
where the creation and annihilation operators satisfy standard commutation relation
\begin{align}
    &[b_{\alpha,\bm k}, b_{\beta,\bm k'}]=[b^\dagger_{\alpha,\bm k}, b^\dagger_{\beta,\bm k'}]=0,\\
  &[b_{\alpha,\bm k}, b^\dagger_{\beta,\bm k'}]=(2\pi)^3\delta_{\alpha,\beta}\delta^{(3)}(\bm k-\bm k'), 
\end{align}
and the polarization tensor \(\epsilon^{\alpha}_{{\mu\nu}}(\bm k)\) satisfies
\begin{align}
  \epsilon^{\alpha}_{{\mu\nu}}(\bm k)=\epsilon^{\alpha}_{\nu\mu}(\bm k),\qquad k^\mu\epsilon^{\alpha}_{{\mu\nu}}(\bm k)=\frac{1}{2}k_\nu\epsilon^{\mu\alpha}_{\mu}(\bm k).
\end{align} 
 There are six independent solutions for the above equations. We can further demand $\epsilon^{\mu\alpha}_{\mu}(\bm k)=0$ and $\epsilon^{\alpha}_{0\mu}(\bm k)=0$. This actually leads to transverse and traceless gauge, and the PF equation still reduces to wave equation, so the expansion with plane waves remains   reasonable. The completeness relation for the polarization tensor  is \cite{vanDam:1970vg}
\bea 
\sum_{\alpha}\epsilon_{\mu\nu}^{*\alpha}(\bm k)\epsilon_{\rho\sigma}^{\alpha'}(\bm k)\delta_{\alpha,\alpha'}=\frac{1}{2}(\bar\eta_{\mu\rho}\bar\eta_{\nu\sigma}+\bar\eta_{\mu\sigma}\bar\eta_{\nu\rho}-\bar\eta_{\mu\nu}\bar\eta_{\rho\sigma})
\eea where 
\bea \bar\eta_{\mu\nu}=\eta_{\mu\nu}-\frac{1}{2}(n_\mu(\bm k)\bar{n}_\nu(\bm k)+n_\nu(\bm k)\bar{n}_\mu(\bm k)).
\eea The vectors $n_\mu(\bm k)$ and $\bar{n}_\mu(\bm k)$ are
\bea
n_\mu(\bm k)=(-1,n_i(\bm k)),\quad \bar{n}_\mu(\bm k)=(1,n_i(\bm k)),\quad n_i(\bm k)=\frac{k_i}{|\bm k|}.
\eea Substituting \eqref{completeY}, the completeness relation becomes 
\bea 
\sum_{\alpha}\epsilon_{\mu\nu}^{*\alpha}(\bm k)\epsilon_{\rho\sigma}^{\alpha'}(\bm k)\delta_{\alpha,\alpha'}=\frac{1}{2}(Y_\mu^AY_{\rho A}Y_\nu^B Y_{\sigma B}+Y_\mu^AY_{\sigma A}Y_\nu^B Y_{\rho B}-Y_\mu^AY_{\nu A}Y_\rho^B Y_{\sigma B}).
\eea In this relation, the arguments of the vector $Y^A_\mu$ are $\Omega_k$ defined in the following \eqref{spherical}.

\subsection{Antipodal matching condition}
In this subsection, we use mode expansion of quantized field to derive the antipodal matching conditions. Starting from \eqref{hmnexp} and using asymptotic expansion of the spherical Bessel function of the first kind
\begin{align}
    j_\ell(\omega r)=\frac{\sin(\omega r-\frac{\pi\ell}{2})}{\omega r}+\frac{\ell(\ell+1)}{2\omega^2r^2}\cos(\omega r-\frac{\pi\ell}{2})+\mathcal{O}\left(\frac{1}{r^3}\right),
\end{align}
we find the large-$r$ expansion of the plane wave 
\begin{align}
  e^{-i\omega t+i\bm k\cdot\bm x}=&4\pi\sum_{\ell m}\frac{i^\ell}{2i\omega r}[e^{-i\omega u-i\pi\ell/2}-e^{-i\omega v+i\pi\ell/2}]Y^*_{\ell, m}(\Omega_k)Y_{\ell, m}(\Omega)\nn\\
  &+4\pi\sum_{\ell m}i^\ell\frac{\ell(\ell+1)}{4\omega^2r^2}[e^{-i\omega u-i\pi\ell/2}+e^{-i(\omega v-\pi\ell/2)}]Y^*_{\ell, m}(\Omega_k)Y_{\ell, m}(\Omega)+\mathcal{O}(r^{-3})
\end{align}
where we have used the spherical coordinates for the spatial position $\bm x$ and wave vector $\bm k$
\bea 
\bm x=(r,\Omega),\quad \bm k=(\omega,\Omega_k).\label{spherical}
\eea 
Therefore, we get the leading order terms at future and past null infinity \footnote{The superscript $+$ is to denote the field at future null infinity and $-$ is to denote the field at past null infinity. }
\begin{align}
  &H_{\mu\nu}^{+{(1)}}(u,\Omega)=\int_0^\infty \frac{d\omega}{\sqrt{4\pi\omega}} \sum_{\ell,m}[c_{{\mu\nu};\omega,\ell,m}e^{-i\omega u}Y_{\ell,m}(\Omega)+\text{h.c.}],\\
  &H_{\mu\nu}^{-{(1)}}(v,\Omega)=\int_0^\infty \frac{d\omega}{\sqrt{4\pi\omega}} \sum_{\ell,m}[\widetilde c_{{\mu\nu};\omega,\ell,m}e^{-i\omega v}Y_{\ell,m}(\Omega)+\text{h.c.}],
\end{align}
where
\begin{align}
  c_{{\mu\nu};\omega,\ell,m}=&\frac{\omega}{(2\pi)^{3/2}i}\int d\Omega_k \sum_{\alpha}\epsilon_{\mu\nu}^{*\alpha}(\bm k)b_{\alpha,\bm k}Y_{\ell,m}^*(\Omega_k),\\
  c_{{\mu\nu};\omega,\ell,m}^\dagger=&\frac{i\omega}{(2\pi)^{3/2}}\int d\Omega_k \sum_{\alpha}\epsilon_{\mu\nu}^{\alpha}(\bm k)b^\dagger_{\alpha,\bm k}Y_{\ell,m}(\Omega_k),\\
  \widetilde c_{{\mu\nu};\omega,\ell,m}=&(-1)^{\ell}\frac{i\omega}{(2\pi)^{3/2}}\int d\Omega_k \sum_{\alpha}\epsilon_{\mu\nu}^{*\alpha}(\bm k)b_{\alpha,\bm k}Y_{\ell,m}^*(\Omega_k),\\
  c_{{\mu\nu};\omega,\ell,m}^\dagger=&(-1)^{\ell}\frac{\omega}{(2\pi)^{3/2}i}\int d\Omega_k \sum_{\alpha}\epsilon_{\mu\nu}^{\alpha}(\bm k)b^\dagger_{\alpha,\bm k}Y_{\ell,m}(\Omega_k).
\end{align} Therefore, the antipodal matching condition for the annihilation and creation operators is 
\bea 
c_{\mu\nu;\omega,\ell,m}=(-1)^{\ell+1}\widetilde{c}_{\mu\nu;\omega,\ell,m},\quad c^\dagger_{\mu\nu;\omega,\ell,m}=(-1)^{\ell+1}\widetilde{c}^\dagger_{\mu\nu;\omega,\ell,m}.
\eea 
Similarly, the subleading terms are
\begin{align}
  &H_{\mu\nu}^{+{(2)}}(u,\Omega)=\int_0^\infty \frac{d\omega}{\sqrt{4\pi\omega}} \sum_{\ell,m}[\frac{i\ell(\ell+1)}{2\omega}c_{{\mu\nu};\omega,\ell,m}e^{-i\omega u}Y_{\ell,m}(\Omega)+\text{h.c.}],\\
  &H_{\mu\nu}^{-{(2)}}(v,\Omega)=\int_0^\infty \frac{d\omega}{\sqrt{4\pi\omega}} \sum_{\ell,m}[\frac{\ell(\ell+1)}{2i\omega}\widetilde c_{{\mu\nu};\omega,\ell,m}e^{-i\omega v}Y_{\ell,m}(\Omega)+\text{h.c.}].
\end{align}

To find antipodal matching condition, we need transform to Fourier space with respect to retarded/advanced time. For leading terms, we find
\begin{align}
  H_{\mu\nu}^{+}(\omega,\Omega)=&\theta(\omega)\sqrt{\frac{\pi}{\omega}}\sum_{\ell,m}c_{{\mu\nu};\omega,\ell,m}Y_{\ell,m}(\Omega)+\theta(-\omega)\sqrt{-\frac{\pi}{\omega}}\sum_{\ell,m}c_{{\mu\nu};-\omega,\ell,m}^\dagger Y^*_{\ell,m}(\Omega),\label{hmn1p}
\end{align}
and similarly
\begin{align}
  H_{\mu\nu}^{-}(\omega,\Omega)
  =&\theta(\omega)\sqrt{\frac{\pi}{\omega}}\sum_{\ell,m}\widetilde{c}_{{\mu\nu};\omega,\ell,m}Y_{\ell,m}(\Omega)+\theta(-\omega)\sqrt{-\frac{\pi}{\omega}}\sum_{\ell,m}\widetilde{c}_{{\mu\nu};-\omega,\ell,m}^\dagger Y^*_{\ell,m}(\Omega)\nn\\
  =&-\theta(\omega)\sqrt{\frac{\pi}{\omega}}\sum_{\ell,m}c_{{\mu\nu};\omega,\ell,m}Y_{\ell,m}(\Omega^P)-\theta(-\omega)\sqrt{-\frac{\pi}{\omega}}\sum_{\ell,m}c_{{\mu\nu};-\omega,\ell,m}^\dagger Y^*_{\ell,m}(\Omega^P),\label{hmn1m}
\end{align} 
where $\Omega^P$ is antipodal to $\Omega=(\theta,\phi)$
\begin{align}
    \Omega^P=(\pi-\theta,\pi+\phi)
\end{align} and the parity transformation of the spherical harmonic function is 
\be 
Y_{\ell,m}(\Omega^P)=(-1)^\ell Y_{\ell,m}(\Omega).
\ee 
Comparing \eqref{hmn1p} and \eqref{hmn1m}, one can find
\begin{align}
    H^{+}_{\mu\nu}(\omega,\Omega)=-H^{-}_{\mu\nu}(\omega,\Omega^P).
\end{align}
To subleading order, we have
\begin{align}
    H^{+(2)}_{\mu\nu}(\omega,\Omega)=H^{-(2)}_{\mu\nu}(\omega,\Omega^P).
\end{align}

\subsubsection*{Electric and magnetic fields}
For linearized gravity, 
we could define electric and magnetic fields analogous to Maxwell theory
\begin{align}
  {\mathcal E}_{mn} = -R_{0m0n} , \qquad {\mathcal B}_{mn} = \frac{1}{2} \, \epsilon_{npq}\, R_{0m}^{\; \; \; \; pq},\qquad m,n=1,2,3
\end{align}
With \eqref{riemann}, we can write them explicitly
\begin{align}
  &{\mathcal E}_{mn} =\frac{1}{2}(\partial_0^2h_{mn}-\partial_0\partial_mh_{0n}-\partial_0\partial_nh_{0m}+\partial_n\partial_mh_{00}),\\
  &{\mathcal B}_{mn} =-\frac{1}{2}\epsilon_{n}{}^{pq}(\partial_p\partial_0h_{mq}-\partial_p\partial_mh_{0q}).
\end{align}
We may expand the electric and magnetic part asymptotically as 
\bea 
\mathcal{E}_{mn}&=&\frac{\mathcal{E}_{mn}^+(u,\Omega)}{r}+\sum_{k=2}^\infty \frac{\mathcal{E}_{mn}^{+(k)}(u,\Omega)}{r^k},\\
\mathcal{B}_{mn}&=&\frac{\mathcal{B}_{mn}^+(u,\Omega)}{r}+\sum_{k=2}^\infty \frac{\mathcal{B}_{mn}^{+(k)}(u,\Omega)}{r^k},\\ 
\mathcal{E}_{mn}&=&\frac{\mathcal{E}_{mn}^-(v,\Omega)}{r}+\sum_{k=2}^\infty \frac{\mathcal{E}_{mn}^{-(k)}(v,\Omega)}{r^k},\\
\mathcal{B}_{mn}&=&\frac{\mathcal{B}_{mn}^-(v,\Omega)}{r}+\sum_{k=2}^\infty \frac{\mathcal{B}_{mn}^{-(k)}(v,\Omega)}{r^k}.
\eea 
Sending to null infinity, we obtain
\begin{align}
  &{\mathcal E}^+_{mn} (u,\Omega)=\frac{1}{2}[\ddot H^+_{mn}(u,\Omega)+n_m(\Omega)\ddot H^+_{0n}(u,\Omega)+n_n(\Omega)\ddot H^+_{0m}(u,\Omega)+n_m(\Omega)n_n(\Omega)\ddot H^+_{00}(u,\Omega)],\\
  &{\mathcal B}^+_{mn} (u,\Omega)=\frac{1}{2}\epsilon_{n}{}^{pq}[n_p(\Omega)\ddot H^+_{mq}(u,\Omega)+n_p(\Omega)n_m(\Omega)\ddot H^+_{0q}(u,\Omega)]
\end{align}
for $\mathcal{I}^+$, and
\begin{align}
  &{\mathcal E}^-_{mn} (v,\Omega)=\frac{1}{2}[\ddot H^-_{mn}(v,\Omega)-n_m(\Omega)\ddot H^-_{0n}(v,\Omega)-n_n(\Omega)\ddot H^-_{0m}(v,\Omega)+n_m(\Omega)n_n(\Omega)\ddot H^-_{00}(v,\Omega)],\\
  &{\mathcal B}^-_{mn} (v,\Omega)=-\frac{1}{2}\epsilon_{n}{}^{pq}[n_p(\Omega)\ddot H^-_{mq}(v,\Omega)-n_p(\Omega)n_m(\Omega)\ddot H^-_{0q}(v,\Omega)]
\end{align}
for $\mathcal{I}^-$. Converting to Fourier space, one find
\begin{align}
  &{\mathcal E}^+_{mn} (\omega,\Omega)=-\frac{1}{2}\omega^2[H^+_{mn}(\omega,\Omega)+n_m(\Omega)H^+_{0n}(\omega,\Omega)+n_n(\Omega)H^+_{0m}(\omega,\Omega)+n_m(\Omega)n_n(\Omega)H^+_{00}(\omega,\Omega)],\\
  &{\mathcal B}^+_{mn} (\omega,\Omega)=-\frac{1}{2}\omega^2\epsilon_{n}{}^{pq}[n_p(\Omega)H^+_{mq}(\omega,\Omega)+n_p(\Omega)n_m(\Omega)H^+_{0q}(\omega,\Omega)],
\end{align}
and
\begin{align}
  &{\mathcal E}^-_{mn}(\omega,\Omega)=-\frac{1}{2}\omega^2[H^-_{mn} (\omega,\Omega)-n_m(\Omega)H^-_{0n} (\omega,\Omega)-n_n(\Omega)H_{0m} (\omega,\Omega)+n_m(\Omega)n_n(\Omega)H^-_{00} (\omega,\Omega)],\\
  &{\mathcal B}^-_{mn}  (\omega,\Omega)=\frac{1}{2}\omega^2\epsilon_{n}{}^{pq}[n_p(\Omega)H^-_{mq} (\omega,\Omega)-n_p(\Omega)n_m(\Omega)H^-_{0q} (\omega,\Omega)].
\end{align}
Using the relation
\begin{align}
  n_i(\Omega^P)=-n_i(\Omega),
\end{align}
we get the antipodal condition for electric and magnetic fields
\begin{align}
  {\mathcal E}^+_{mn}(\omega,\Omega)=-{\mathcal E}^-_{mn}(\omega,\Omega^P),\qquad {\mathcal B}^+_{mn}(\omega,\Omega)=-{\mathcal B}^-_{mn}(\omega,\Omega^P).
\end{align}

As a matter of fact, we could discuss the antipodal matching conditions for linearized Riemann tensor which is more general than electric and magnetic fields. We expand the Riemann tensor asymptotically as 
\bea 
R_{\mu\nu\rho\sigma}&=&\frac{R^+_{\mu\nu\rho\sigma}(u,\Omega)}{r}+\sum_{k=2}^\infty \frac{R^{+(k)}_{\mu\nu\rho\sigma}(u,\Omega)}{r^k},\\ 
R_{\mu\nu\rho\sigma}&=&\frac{R^-_{\mu\nu\rho\sigma}(v,\Omega)}{r}+\sum_{k=2}^\infty \frac{R^{+(k)}_{\mu\nu\rho\sigma}(v,\Omega)}{r^k}.
\eea At future/past null infinity, we find 
\begin{align}
    R^+_{\mu\nu\rho\sigma}(u,\Omega)=&\frac{1}{2}\left(n_\rho n_\nu \ddot{H}^+_{\rho\sigma}(u,\Omega)-n_\rho n_\mu \ddot{H}^+_{\nu\sigma}(u,\Omega)-n_\sigma n_\nu \ddot{H}^+_{\mu\rho}(u,\Omega)+n_\rho n_\sigma \ddot{H}^+_{\nu\rho}(u,\Omega)\right),\\ 
R^-_{\mu\nu\rho\sigma}(v,\Omega)=&\frac{1}{2}\left(n_\rho n_\nu \ddot{H}^-_{\rho\sigma}(v,\Omega)-n_\rho n_\mu\ddot{H}^-_{\nu\sigma}(v,\Omega)-n_\sigma n_\nu \ddot{H}^-_{\mu\rho}(v,\Omega)+n_\rho n_\sigma \ddot{H}^-_{\nu\rho}(v,\Omega)\right).
\end{align} 
In Fourier space, they are equivalent to 
\bea 
R^+_{\mu\nu\rho\sigma}(\omega,\Omega)&=&-\frac{1}{2}\omega^2\left(n_\rho n_\nu H^+_{\rho\sigma}(\omega,\Omega)-n_\rho n_\mu H^+_{\nu\sigma}(\omega,\Omega)-n_\sigma n_\nu H^+_{\mu\rho}(\omega,\Omega)+n_\rho n_\sigma H^+_{\nu\rho}(\omega,\Omega)\right),\nn\\
R^-_{\mu\nu\rho\sigma}(\omega,\Omega)&=&-\frac{1}{2}\omega^2\left(n_\rho n_\nu H^-_{\rho\sigma}(\omega,\Omega)-n_\rho n_\mu H^-_{\nu\sigma}(\omega,\Omega)-n_\sigma n_\nu H^-_{\mu\rho}(\omega,\Omega)+n_\rho n_\sigma H^-_{\nu\rho}(\omega,\Omega)\right).\nn
\eea Consequently, we find the antipodal condition 
\bea 
R^+_{\mu\nu\rho\sigma}(\omega,\Omega)=-R^-_{\mu\nu\rho\sigma}(\omega,\Omega^P).
\eea 

\subsection{Canonical quantization}\label{canoquant}
In this subsection, we use mode expansion of quantized field to compute the fundamental commutator of shear tensor in the transverse and traceless gauge. The result will be same to the one in Bondi gauge.

Starting from \eqref{hmnexp}, switching to retarded frame, and approaching future null infinity, we find
\begin{align}
  h_{AB}(t,\bm x)=&\sum_{\alpha}\int \frac{d^3\bm k}{(2\pi)^3}\frac{1}{\sqrt{2\omega_{\bm k}}}r^2Y^i_AY^j_B[\epsilon^{*\alpha}_{ij}(\bm k)b_{\alpha,\bm k}e^{-i\omega t+i\bm k\cdot\bm x}+\epsilon^{\alpha}_{ij}(\bm k)b^\dagger_{\alpha,\bm k}e^{i\omega t-i\bm k\cdot\bm x}]\nn\\
  =&r\int_0^\infty \frac{d\omega}{\sqrt{4\pi\omega}}\sum_{\ell m}[c_{i,j;\omega,\ell,m}Y^i_AY^j_BY_{\ell,m}(\Omega)e^{-i\omega u}+c^\dagger_{i,j;\omega,\ell,m}Y^i_AY^j_BY_{\ell,m}^*(\Omega)e^{i\omega u}]+\mathcal{O}(1),
\end{align}
where the boundary creation and annihilation operators are
\begin{align}
  c_{i,j;\omega,\ell,m}=&\frac{\omega}{(2\pi)^{3/2}i}\int d\Omega_k \sum_{\alpha}\epsilon_{ij}^{*\alpha}(\bm k)b_{\alpha,\bm k} Y_{\ell,m}^*(\Omega_k), \label{cannihilation2}\\
  c^\dagger_{i,j;\omega,\ell,m}=&\frac{i\omega}{(2\pi)^{3/2}}\int d\Omega_k \sum_{\alpha}\epsilon_{ij}^{\alpha}(\bm k)b^\dagger_{\alpha,\bm k} Y_{\ell,m}(\Omega_k).\label{ccreation2}
\end{align}
One can insert back the coefficient $\sqrt{32\pi G}$ and  read out the shear tensor
\begin{align}
  C_{AB}(u,\Omega)=\sqrt{32\pi G}\int_0^\infty \frac{d\omega}{\sqrt{4\pi\omega}}\sum_{\ell m}[c_{i,j;\omega,\ell,m}Y^i_AY^j_BY_{\ell,m}(\Omega)e^{-i\omega u}+c^\dagger_{i,j;\omega,\ell,m}Y^i_AY^j_BY_{\ell,m}^*(\Omega)e^{i\omega u}].
\end{align}
From the completeness relation,
we find
\begin{align}
  [c_{i,j;\omega,\ell,m},c_{i',j';\omega',\ell',m'}]=&[c^\dagger_{i,j;\omega,\ell,m},c^\dagger_{i',j';\omega',\ell',m'}]=0,\\
  [c_{i,j;\omega,\ell,m},c^\dagger_{i',j';\omega',\ell',m,}]=&\frac{1}{2}\delta(\omega-\omega')\int d\Omega \big(\bar{\eta}_{i,i'}\bar{\eta}_{j,j'}-\bar{\eta}_{i,j}\bar{\eta}_{i',j'}+\bar{\eta}_{i,j'}\bar{\eta}_{i',j}\big) Y_{\ell,m}^*(\Omega)Y_{\ell',m'}(\Omega).
\end{align} where 
\bea 
\bar{\eta}_{ij}=\delta_{ij}-n_i n_j=Y_i^A(\Omega)Y_{jA}(\Omega).
\eea 
Thus we can obtain the following commutator
\begin{align}
  [C_{AB}(u,\Omega),C_{CD}(u',\Omega')]=8\pi G i P_{ACDB}\alpha(u-u')\delta(\Omega-\Omega'),
\end{align}
as expected.

\section{Commutators}\label{commutator}
In this appendix, we will provide some details on the calculation of commutators among flux operators. We will take $[\mathcal{M}_Y,\mathcal{M}_Z]$ as an example. To simplify computation, we start from rewriting $\mathcal{M}_Y$ (without normal order written out due to its irrelevance to non-central terms)
\begin{align}
  \mathcal{M}_Y=&\frac{1}{32\pi G}\int du d\Omega Y^A(u,\Omega)(\dot{C}^{BC}\nabla^DC^{EF}-C^{BC}\nabla^D\dot{C}^{EF})P_{ABCDEF}\nonumber\\
  =&\frac{1}{32\pi G}\int du d\Omega \dot C^{BC}\Delta_{BC}(Y;C;u,\Omega),\label{gravsuperrotflux}
\end{align}
where $\Delta_{BC}(Y;C;u,\Omega)$ is given by \eqref{Deltaab}. Then we can compute straightforwardly
\begin{align}
  [\mathcal{M}_Y,\mathcal{M}_Z]=&\frac{1}{32\pi G}\int du d\Omega [\mathcal{M}_Y,\dot C^{BC}\Delta_{BC}(Z;C;u,\Omega)]\nonumber\\
  =&\frac{-i}{32\pi G}\int du d\Omega\Big[\big[\Delta^{BC}(Y;\dot C;u,\Omega)+\frac{1}2\Delta^{BC}(\dot Y;C;u,\Omega)\big]\Delta_{BC}(Z;C;u,\Omega)\\
  &+2\dot C^{BC}Z^A\nabla^D\Delta^{EF}(Y;C;u,\Omega)\rho_{AEFDBC}+\dot C^{BC}\nabla^DZ^A\Delta^{EF}(Y;C;u,\Omega)P_{AEFDBC}\Big]\nonumber\\
  &+\frac{i}{64\pi G}\int du d\Omega du'\alpha(u'-u)\dot C^{BC}\nn\\
  &\times\big[2\rho_{AEFDBC}Z^A\nabla^D\Delta^{EF}(\dot Y;C;u',\Omega)+P_{AEFDBC}\nabla^DZ^A\Delta^{EF}(\dot Y;C;u',\Omega)\big].\nonumber
\end{align}
Using the integration by part several times, we can obtain
\begin{align}
  [\mathcal{M}_Y,\mathcal{M}_Z]=&\frac{-i}{32\pi G}\int du d\Omega\big[\Delta^{EF}(Y;\dot C)\Delta_{EF}(Z;C)-\Delta^{EF}(Y;C)\Delta_{EF}(Z;\dot C)\big]\nonumber\\
  &+\frac{i}{64\pi G}\int du d\Omega du'\alpha(u'-u)\Delta_{EF}(\dot Z;C;u,\Omega)\Delta^{EF}(\dot Y;C;u',\Omega).
\end{align}
The non-local term is precisely the previous $N_M(Y,Z)$, while for local terms, one can further simplify to get
\begin{align}
  \frac{i}{32\pi G}\int du d\Omega\dot C_{EF}\big[\Delta^{EF}(Y;\Delta(Z;C))-\Delta^{EF}(Z;\Delta(Y;C))\big].
\end{align}

To form the local operators, we need use an identity
\begin{align}
  \Delta_{EF}(Y;\Delta(Z;C))-\Delta_{EF}(Z;\Delta(Y;C))=\Delta_{EF}([Y,Z];C)+2o(Y,Z)C^{BC}Q_{EFBC},
\end{align} 
whose proof demand some properties of higher rank tensor. The main ones are
\begin{align}
  2\rho_{ABCDEF}P^{GIJHBC}=\gamma_{AD}P^{GIJH}{}_{EF},\qquad 2\rho_{AIJDEF}P^{GIJHBC}=\gamma_{AD}P^{GEFH}{}_{BC},
\end{align}
\begin{align}
  P_{ABCDEF}P^{GIJHBC}-P^G{}_{BC}{}^H{}_{EF}P_A{}^{IJ}{}_D{}^{BC}=0,
\end{align}
and
\begin{align}
  2Q^{ABCD}Q_{EF}{}^{IJ}=\gamma^{CB}P^{DIJA}{}_{EF}-\gamma^{AD}P^{BIJC}{}_{EF}.
\end{align}
Now it is easy to find the local parts of $[\mathcal{M}_Y,\mathcal{M}_Z]$ can be written as
\begin{align}
  &\frac{i}{32\pi G}\int du d\Omega\dot C^{EF}\big[\Delta_{EF}([Y,Z];C)+2o(Y,Z)C^{IJ}Q_{EFIJ}\big]=i\mathcal{M}_{[Y,Z]}+2i\mathcal{O}_{o(Y,Z)}.
\end{align}
As for central charges, one need start from correlation functions of shear tensor, and we will not show the details here.

For other commutators, we provide the key identities that may be used. The following identity is useful for the calculation of $[\mathcal{T}_f,\mathcal{M}_Y]$
\begin{align}
  P_{ABCDEF}+\epsilon_{DA}Q_{EFBC}-\frac{1}{4}\gamma_{AD}P_{BEFC}=0.
\end{align}
To calculate $[\mathcal{M}_Y,\mathcal{O}_g]$, one might make use of
\begin{align}
  2Q^{BC}{}_{ EF}\rho_{ABCDGH}=\gamma_{AD}Q_{GHEF},
\end{align}
and
\begin{align}
  P_{ABCD}{}^{EF}Q_{EFIJ}-P_{AIJD}{}^{EF}Q_{EFBC}=\gamma_{AD}Q_{BCIJ}.
\end{align}
The remaining commutators are relatively straightforward.

\section{Conserved current for duality transformation}\label{dualcur}
The PF action is not invariant under duality transformation. Just like the electromagnetic theory, we may construct a duality symmetric action 
\bea 
S[h,\widetilde{h}]=\frac{1}{2}(S_{\rm PF}[h]+S_{\rm PF}[\widetilde{h}]).
\eea 
One can derive the equations of motion from this symmetric action. More importantly, the action is invariant under duality transformation. To prove this, we note that the infinitesimal duality transformation is 
\bea 
\delta_\epsilon h_{\mu\nu}=\epsilon\widetilde{h}_{\mu\nu},\qquad \delta_{\epsilon}\widetilde{h}_{\mu\nu}=-\epsilon h_{\mu\nu}.
\eea Therefore, the variation of the symmetric action is 
\bea 
\delta_\epsilon S[h,\widetilde{h}]&=&-\frac{1}{64\pi G}\int d^4 x L^{\mu_1\mu_2\cdots\mu_6}\partial_{\mu_1}h_{\mu_2\mu_3}\delta_\epsilon \partial_{\mu_4}h_{\mu_5\mu_6}-\frac{1}{64\pi G}\int d^4 x L^{\mu_1\mu_2\cdots\mu_6}\partial_{\mu_1}\widetilde{h}_{\mu_2\mu_3}\delta_\epsilon \partial_{\mu_4}\widetilde{h}_{\mu_5\mu_6}\nn\\&=&-\frac{\epsilon}{64\pi G}\int d^4 x L^{\mu_1\mu_2\cdots\mu_6}\partial_{\mu_1}h_{\mu_2\mu_3}\partial_{\mu_4}\widetilde{h}_{\mu_5\mu_6}+\frac{\epsilon}{64\pi G}\int d^4 x L^{\mu_1\mu_2\cdots\mu_6}\partial_{\mu_1}\widetilde{h}_{\mu_2\mu_3}\partial_{\mu_4}h_{\mu_5\mu_6}\nn\\&=&0.
\eea 
At the last step, we used the fact that the tensor $L^{\mu_1\mu_2\cdots\mu_6}$ is invariant under the exchange of indices 
\bea 
\mu_1\mu_2\mu_3\leftrightarrow \mu_4\mu_5\mu_6.
\eea Now using the Noether's theorem, the conserved current is 
\bea 
j_{\text{duality}}^\mu&=&\frac{1}{2}\frac{\partial \mathcal{L}_{\rm PF}(h)}{\partial (\partial_\mu h_{\rho\sigma})}\delta h_{\rho\sigma}+\frac{1}{2}\frac{\partial \mathcal{L}_{\rm PF}(\widetilde{h})}{\partial (\partial_\mu \widetilde{h}_{\rho\sigma})}\delta \widetilde{h}_{\rho\sigma}\nn\\&=&\frac{1}{64\pi G}L^{\mu\rho\sigma\mu_4\mu_5\mu_6}(h_{\rho\sigma}\partial_{\mu_4}\widetilde{h}_{\mu_5\mu_6}-\widetilde{h}_{\rho\sigma}\partial_{\mu_4}h_{\mu_5\mu_6}).
\eea At the first line, $\mathcal{L}_{\rm PF}(h)$ is the Lagrangian density 
\bea 
\mathcal{L}_{\rm PF}(h)=-\frac{1}{64\pi G}L^{\mu_1\mu_2\cdots\mu_6}\partial_{\mu_1}h_{\mu_2\mu_3}\partial_{\mu_4}h_{\mu_5\mu_6}.
\eea At the second line, we have discarded the constant $\epsilon$. One can use the equations of motion to prove the conservation 
\bea 
\partial_\mu j_{\text{duality}}^\mu=0.
\eea

\bibliography{refs}

\end{document}